\documentclass[sn-mathphys-num]{sn-jnl}

\usepackage{graphicx}
\usepackage{subfigure}
\usepackage{epstopdf}
\usepackage{multirow}
\usepackage{amsmath,amssymb,amsfonts}
\usepackage{amsthm}
\usepackage{mathrsfs}
\usepackage[title]{appendix}
\usepackage{xcolor}
\usepackage{textcomp}
\usepackage{manyfoot}
\usepackage{booktabs}
\usepackage{algorithm}
\usepackage{algorithmicx}
\usepackage{algpseudocode}
\usepackage{listings}
\usepackage{slashed}
\usepackage[compat=1.1.0]{tikz-feynman} 
\tikzfeynmanset{warn luatex=false} 

\begin{document}
	
	\title[Decoding Heavy Top-Philic Resonances at the HL-LHC]{Decoding Heavy Top-Philic Resonances at the HL-LHC: From Parton Kinematics to Deep Learning Signatures}
	
	\author*[1]{\fnm{H.} \sur{Boukhrouf}}\email{boukhrouf.hadjer@univ-jijel.dz}
	\author[1]{\fnm{Z.} \sur{Belghobsi}}\email{belghobsi.zouina@univ-jijel.dz}
	
	\affil[1]{\orgdiv{Laboratory of Theoretical Physics (LPTh), Department of Physics, Faculty of Exact Sciences and Informatics}, \orgname{University of Jijel}, \orgaddress{\street{BP 98, Ouled Aissa}, \postcode{18000} \city{Jijel}, \country{Algeria}}}
	
	\abstract{
		The stabilization of the electroweak scale strongly motivates the search for new heavy resonances and top-partner states. We investigate the discovery potential of exotic resonances mediating the production of vector-like quarks at the High-Luminosity LHC ($\sqrt{s} = 13.6$ TeV), focusing on the $pp \to X \to t \bar{t}_p$ cascade decay. Using a model-independent Effective Field Theory approach, we classify the intermediate mediator $X$ by its spin (0 and 1) and color (singlet and octet) representations. Parton-level kinematics demonstrate that the normalized differential cross-section with respect to the transverse momentum provides a robust observable for spin discrimination. To address the experimental challenges of highly boosted, semi-resolved hadronic decays, we introduce a cut-flow strategy based on an inclusive four-jet invariant mass reconstruction, $M(4j)$. Because this standard approach remains sensitive to systematic uncertainties, we implement a Generalized BSM Tagger based on Deep Neural Networks (DNNs). By exploiting non-linear multi-jet correlations, the DNN achieves robust background rejection factors ranging from $\mathcal{O}(600)$ to $\mathcal{O}(1000)$ while preserving the underlying partonic signatures. Projecting to an integrated luminosity of $3000\text{ fb}^{-1}$, this combined strategy yields a projected statistical significance of $3.74\sigma$ for the dominant color-octet vector channel, establishing strong evidence potential and offering a robust phenomenological baseline for future HL-LHC searches.
	}
	
	
	\maketitle
	
	\section{Introduction}\label{sec1}
	
	Despite the success of the Standard Model (SM) of particle physics \cite{Glashow:1961, Weinberg:1967, Salam:1968}, the hierarchy problem remains a major theoretical challenge \cite{Susskind:1979, tHooft:1979}. The large energy gap between the electroweak scale, where the Higgs vacuum expectation value is $v \sim 246$~GeV, and the Planck scale, $M_{Pl} \sim 10^{19}$~GeV, requires extreme fine-tuning. Specifically, the physical Higgs boson \cite{ATLAS:2012_Higgs, CMS:2012_Higgs} mass receives quadratically divergent radiative corrections from heavy particle loops. The top quark loop provides the largest ultraviolet-sensitive contribution, suggesting the presence of new physics at the TeV scale to naturalize the electroweak sector \cite{Giudice:2008_NaturallySpeaking, ArkaniHamed:1998_ADD, Randall:1999_RS}.
	
	To stabilize the Higgs mass, several Beyond Standard Model (BSM) frameworks, such as Composite Higgs with partial compositeness and Little Higgs models \cite{Kaplan:1984, MinimalCompositeHiggs, Contino:2010_CompositeHiggs}, which introduce vector-like fermions. These states, often called Vector-Like Quarks (VLQs) or heavy top partners ($t_p$) carry color charge and transform vectorially under the SM electroweak gauge group \cite{TopPartnerGuide}. Unlike chiral SM quarks, VLQs acquire gauge-invariant bare masses at the multi-TeV scale without relying on the Higgs mechanism, allowing them to evade electroweak precision constraints. These top partners generate loop diagrams that cancel the quadratic divergences induced by the SM top quark \cite{Buchkremer:2013, ExoticDecays}. Searching for VLQs is therefore a primary goal of the physics program at the Large Hadron Collider (LHC) \cite{ATLAS:VLQ_Combine, CMS:VLQ_Search, ATLAS:2018_VLQ_Hadronic, CMS:2020_FourTops_Run2, CMS:2018_FourTops, CMS:2013_VLQ}. However, identifying the properties of the heavy resonances that mediate VLQ production remains challenging. In many BSM extensions, top partners are produced via the $s$-channel exchange of a heavy intermediate resonance $X$, decaying into a third-generation SM quark and a VLQ ($pp \to X \to t \bar{t}_p$). The event kinematics depend heavily on the spin of this mediator, which can be a CP-even scalar ($J^P = 0^+$), a CP-odd pseudoscalar ($J^P = 0^-$), or a vector ($J^P = 1^-$).\\
	
	The resonance dynamics are also determined by the color representation of the mediator, which can be a QCD color-singlet ($\mathbf{1}$) or color-octet ($\mathbf{8}$). Color-singlet states generally have lower production rates at hadron colliders. In contrast, color-octet resonances couple directly to initial-state gluons, resulting in larger cross-sections and different color-flow topologies \cite{Chivukula:2011_Coloron, Haisch:2011_ColorOctet, Dutta:2013_TopOctet, Kilic:2010_Vectorlike}. Recent work on hypercolor models has highlighted the phenomenological interest of embedding top-partner systems in such exotic color representations \cite{Cacciapaglia:2026_ExoticColour, Cacciapaglia:2021_UnusualTopPartners}.\\
	
	To analyze these diverse spin and color signatures independently of a specific ultraviolet completion, we use an Effective Field Theory (EFT) framework \cite{Skiba:2010_EFT, Degrande:2012_EFT}. The effective Lagrangian describes the interactions between the mediator $X$ and the SM-VLQ sector using specific operators \cite{Darme:2021_TopPhilic}. Scalar and pseudoscalar mediators ($J^P = 0^{\pm}$) couple through Yukawa-like terms associated with chiral symmetry breaking \cite{Cacciapaglia:2015_Octet}, while vector mediators ($J^P = 1^-$) interact via covariant currents indicating extended gauge symmetries \cite{Ferretti:2016_Gauge, Langacker:2008_Zprime}. Evaluating these operators in both color-singlet and color-octet configurations allows us to study how Lorentz structures and QCD color flow affect the final-state kinematics \cite{Chivukula:2011_Coloron, Darme:2018_Sgluons}.
	
	EFT methods are widely used to parameterize deviations in the top-quark sector, supported by global fits and automated tools for Standard Model Effective Field Theory (SMEFT) \cite{Barducci:2018_SMEFT_Top, Degrande:2020_SMEFTatNLO, Durieux:2019_SMEFT_Top}. Following this approach, we utilize a generalized BSM EFT framework. While SMEFT typically focuses on dimension-six operators that modify SM interactions, our setup includes heavy dynamical mediators and describes their loop-induced production through dimension-five operators.
	
	This work investigates the production of heavy top-philic resonances that decay into a top quark and a vector-like partner ($pp \to X \to t \bar{t}_p$). Since the VLQ subsequently decays into a top quark and a heavy electroweak or Higgs boson ($t_p \to tW/tZ/tH$), the characteristic final state features a top-quark pair accompanied by bosonic decay products. The primary challenge in isolating this signal is the overwhelming Standard Model $t\bar{t}$+jets background. Moreover, recent limits from the CMS and ATLAS collaborations \cite{CMS:2025_VLQ_Review, ATLAS:2025_VLQ_Combination, CMS:2024_VLQ_Hadronic} have pushed the lower bounds on the $t_p$ mass to approximately $1.5$~TeV. At these multi-TeV scales, the decay products are highly boosted, necessitating a dedicated focus on the all-hadronic decay regime to maximize signal acceptance \cite{Darme:2025_BoostingBeyond}. While our current phenomenological analysis establishes a robust leading-order baseline, fully exploiting the future High-Luminosity LHC dataset will eventually require incorporating Next-to-Leading Order (NLO) QCD corrections and exploring single-production topologies \cite{Frederix:2018_FourTopsNLO, Deandrea:2018_NLOSingleVLQ} as natural extensions of this framework.
	
	From an experimental perspective, isolating this specific BSM signal from the overwhelming SM background poses a formidable challenge. Standard cut-based analyses, which rely on rigid kinematic thresholds, often exhibit limited sensitivity to highly boosted multi-jet topologies. Such traditional methods frequently suffer from out-of-cone energy losses and fail to fully exploit the non-linear, multi-dimensional correlations inherent to the final-state kinematics. To address these experimental bottlenecks at the HL-LHC operating at a center-of-mass energy of 13.6 TeV, we supplement our kinematic cut-flow with a Deep Learning framework. Recent advancements in collider phenomenology have demonstrated the remarkable efficacy of machine learning, particularly in jet substructure and signal discrimination\cite{Guest:2018yhq,Larkoski:2017jix,Kasieczka:2019pej}. Building upon this progress, we implement a Generalized BSM Tagger based on Deep Neural Networks (DNNs). By leveraging multi-jet observables, this DNN architecture efficiently learns complex phase-space correlations. This approach not only recovers underlying kinematic features but also drastically improves background rejection, providing a highly robust strategy for future HL-LHC searches.
	
	\section{Theoretical Framework and Effective Field Theory}\label{sec2}
	
	We use a model-independent Effective Field Theory (EFT) framework \cite{Gripaios:2015_EFT, Dermisek:2024_2HDMEFT} to study heavy top-philic resonances. We extend the Standard Model (SM) particle spectrum by introducing a heavy Vector-Like Quark (VLQ) acting as a top partner ($t_p$), alongside a multi-TeV intermediate resonance ($X$). Since the VLQ transforms vectorially under the SM electroweak gauge group, its bare mass is gauge-invariant, which allows it to evade strict electroweak precision constraints \cite{Deandrea:2018_VLQDoublets}.
	
	The primary phenomenological signature is the s-channel production topology $pp \to X \to t\bar{t}_p$ (and its charge conjugate). To cover a wide range of possible mediator properties, we classify $X$ by its spin (0 and 1) and QCD color (Color-Singlet and Color-Octet) representations. Gauge invariance and the presumed underlying dynamics determine the mass dimension of the operators driving the production of $X$. Moreover, using higher-dimensional operators for third-generation quarks introduces QCD mixing effects. The systematic matching of these effective operators at Next-to-Leading Order (NLO) in QCD \cite{Zhang:2014_NLO_EFT} justifies using this EFT approach to evaluate top-philic cascade decays.
	
	\subsection{The Spin-0 Sector: Loop-Induced Dynamics}
	
	For the spin-0 sector, we consider both CP-even (scalar, $S$) and CP-odd (pseudoscalar, $P$) states. In composite Higgs frameworks, these scalars are often pseudo-Nambu-Goldstone bosons (pNGBs) protected by a shift symmetry. Gauge invariance prevents direct tree-level minimal couplings between these spin-0 states and SM gauge bosons. Therefore, the effective couplings are formulated through dimension-five operators. Following standard EFT treatments for heavy composite scalars and Axion-Like Particles (ALPs) \cite{Bauer:2017_ALPs, Brivio:2017_ALPs}, the leading-order interaction with QCD gluons involves the field strength tensor ($\Phi G_{\mu\nu} G^{\mu\nu}$). Their production via gluon-gluon fusion ($gg \to X$) is loop-induced and parameterized by an effective coupling $g_{\text{eff}}$ and a UV cutoff $\Lambda$.
	
	For the color-singlet states ($S_1, P_1$), the effective Lagrangians for the mass terms, gluon fusion, and Yukawa interactions are:
	\begin{equation}
		\mathcal{L}_{S_1} = \frac{1}{2} (\partial_\mu S_1)(\partial^\mu S_1) - \frac{1}{2} M_{S_1}^2 S_1^2 + \frac{g_{\text{eff}}}{4\Lambda} S_1 G_{\mu\nu}^a G^{a\mu\nu} - y_{S_1} S_1 \left( \bar{t}_p t + \bar{t} t_p \right) \,,
	\end{equation}
	\begin{equation}
		\mathcal{L}_{P_1} = \frac{1}{2} (\partial_\mu P_1)(\partial^\mu P_1) - \frac{1}{2} M_{P_1}^2 P_1^2 + \frac{g_{\text{eff}}}{8\Lambda} P_1 G_{\mu\nu}^a \tilde{G}^{a\mu\nu} + \left[ i y_{P_1} P_1 \left( \bar{t}_p \gamma_5 t \right) + \text{h.c.} \right] \,,
	\end{equation}
	where $G_{\mu\nu}^a$ denotes the gluon field strength tensor, $\tilde{G}^{a\mu\nu}$ is its dual, and $g_{\text{eff}}$ parameterizes the effective coupling constant.
	
	For color-octet mediators ($\mathbf{8}$ of $SU(3)_C$), the fields carry color indices ($a=1,\dots,8$) and couple via the covariant derivative $D_\mu = \partial_\mu - i g_s T^a A_\mu^a$. Contracting the dimension-five operators with the symmetric $SU(3)$ structure constants $d^{abc}$, the color-octet Lagrangians ($S_8, P_8$) become:
	\begin{equation}
		\mathcal{L}_{S_8} = \frac{1}{2} (D_\mu S_8^a)(D^\mu S_8^a) - \frac{1}{2} M_{S_8}^2 S_8^a S_8^a + \frac{g_{\text{eff}}}{4\Lambda} d^{abc} S_8^a G_{\mu\nu}^b G^{c\mu\nu} + \left[ y_{S_8} S_8^a \left( \bar{t}_p T^a t \right) + \text{h.c.} \right] \,
	\end{equation}
	\begin{equation}
		\mathcal{L}_{P_8} = \frac{1}{2} (D_\mu P_8^a)(D^\mu P_8^a) - \frac{1}{2} M_{P_8}^2 P_8^a P_8^a + \frac{g_{\text{eff}}}{8\Lambda} d^{abc} P_8^a G_{\mu\nu}^b \tilde{G}^{c\mu\nu} + \left[ i y_{P_8} P_8^a \left( \bar{t}_p T^a \gamma_5 t \right) + \text{h.c.} \right] \,.
	\end{equation}
	This color-octet structure modifies the color flow and final-state radiation compared to the singlet equivalents, which impacts multi-top final state topologies.
	
	\subsection{The Spin-1 Sector: Hidden Local Symmetry and Renormalizability}
	We describe the spin-1 mediators ($V$) as heavy composite gauge bosons using Hidden Local Symmetry (HLS) \cite{Bando:1988_HLS, Casalbuoni:1985_BESS, Pappadopulo:2014_HVT}. Unlike the scalar sector, this framework allows direct, tree-level minimal couplings to SM fermion currents. Because the interaction between the vector mediator and the fermion bilinear ($\bar{\psi}\gamma^\mu\psi$) forms a dimension-four operator, the EFT expansion is dominated by these renormalizable terms \cite{Skiba:2010_EFT}.
	
	For the color-singlet vector ($V_1$), acting as a heavy $Z^\prime$ boson \cite{Davighi:2021_AnomalousZ}, the interactions use standard chiral projections $P_{L/R} = \frac{1}{2}(1 \pm \gamma_5)$:
	\begin{equation}
		\mathcal{L}_{V_1} = -\frac{1}{4} V_{\mu\nu} V^{\mu\nu} + \frac{1}{2} M_{V_1}^2 V_\mu V^\mu + \left[ V_\mu \bar{t}_p \gamma^\mu (g_{1L} P_L + g_{1R} P_R) t + \text{h.c.} \right] \,.
	\end{equation}
	
	For the color-octet vector ($V_8$), often modeled as a massive coloron, the interaction with the vector-like color current is:
	\begin{equation}
		\mathcal{L}_{V_8} = -\frac{1}{4} V_{\mu\nu}^a V^{\mu\nu, a} + \frac{1}{2} M_{V_8}^2 V_\mu^a V^{\mu,a} + \left[ V_\mu^a \bar{t}_p T^a \gamma^\mu (g_{8L} P_L + g_{8R} P_R) t + \text{h.c.} \right] \,.
	\end{equation}
	The production mechanisms for $V_1$ and $V_8$ differ significantly. The Landau-Yang theorem forbids a massive spin-1 particle from decaying into or being produced by two massless photons, but this restriction does not apply to QCD because gluons carry color charge \cite{Cacciari:2015_LandauYang}. As a result, $V_8$ couples directly to initial-state gluons via three-point vertices, yielding much larger hadronic cross-sections than $V_1$, which is primarily produced via $q\bar{q}$ annihilation.
	
	\vspace{0.5cm}
	\noindent\textbf{EFT Validity:} \\
	We must ensure the validity of the EFT expansion, particularly for the dimension-five gluon-fusion operators. These terms are valid only when the partonic center-of-mass energy ($\sqrt{\hat{s}}$) is below the new physics scale ($\Lambda$) \cite{Contino:2016_EFT_Validity}. For a resonance mass of $M_X \approx 3$~TeV, the relevant scattering scale is $\sqrt{\hat{s}} \approx 3$~TeV. Assuming a typical composite Higgs cutoff scale of $\Lambda = 10$~TeV \cite{Panico:2015_CompositeHiggs}, the condition $\sqrt{\hat{s}} \ll \Lambda$ is satisfied. This avoids unphysical unitarity violations and maintains perturbative control.
	
	\subsection{Vector-Like Top Partner Decay Topology}
	After production, the heavy top partner $t_p$ decays into SM third-generation particles. The effective Lagrangian for interactions with electroweak gauge bosons and the Higgs boson follows standard VLQ formalisms \cite{TopPartnerGuide, Buchkremer:2013}:
	\begin{equation}
		\begin{split}
			\mathcal{L}_{t_p \to \text{SM}} &= \kappa_{t_p} \sum_{i=1}^{3} \bigg[ \frac{g_w}{\sqrt{2}} \bar{t}_p \gamma^\mu (c_L^{W} P_L + c_R^{W} P_R) q_{d_i} W_\mu^+ \\
			&\quad + \frac{g_w}{2 c_w} \bar{t}_p \gamma^\mu (c_L^{Z} P_L + c_R^{Z} P_R) q_{u_i} Z_\mu 
			- \frac{M_{t_p}}{v} \bar{t}_p (c_L^{H} P_L + c_R^{H} P_R) q_{u_i} H \bigg] + \text{h.c.}
		\end{split}
	\end{equation}
	At high center-of-mass energies ($M_{t_p} \gg M_{W,Z}$), the Goldstone Boson Equivalence Theorem \cite{Cornwall:1974_GBET, Lee:1977_GBET} predicts democratic branching ratios (approximately $2:1:1$ for $W:Z:H$ respectively).
	
\begin{figure}[htbp]
		\centering
		\begin{minipage}{0.47\textwidth}
			\centering
			\begin{tikzpicture}
				\begin{feynman}
					\vertex (q1) at (0, 1.5) {$q$};
					\vertex (q2) at (0, -1.5) {$\bar{q}$};
					\vertex [dot, label=above:{$g_{q}$}] (v1) at (1, 0) {};
					\vertex [dot, label=above:{$g_{V}$}] (v2) at (3.7, 0) {};
					\vertex [dot, label=below:{$\kappa_{t_p}$}] (v3) at (5, -1) {};
					\vertex (t)  at (6, 1.5) {$t$};
					\vertex (w)  at (7, -0.4) {$W^- / Z / H$};
					\vertex (b)  at (7, -2.4) {$\bar{b} / \bar{t}$};
					\diagram* {
						(q1) -- [fermion] (v1),
						(q2) -- [anti fermion] (v1),
						(v1) -- [boson, edge label=$V_8$, edge label'=$V_1$] (v2),
						(v2) -- [fermion] (t),
						(v2) -- [fermion, edge label=$\bar{t}_p$] (v3),
						(v3) -- [boson] (w),
						(v3) -- [anti fermion] (b)
					};
				\end{feynman}
			\end{tikzpicture}
			{(a) Quark-Antiquark Annihilation}
		\end{minipage}\hfill
		\begin{minipage}{0.47\textwidth}
			\centering
			\begin{tikzpicture}
				\begin{feynman}
					\vertex (g1) at (0, 1.6) {$g$};
					\vertex (g2) at (0, -1.6) {$g$};
					\vertex [dot, label=above:{$\frac{g_{\text{eff}}}{\Lambda}$}] (v1) at (2, 0) {};
					\vertex [dot, label=above:{$y_{X}$}] (v2) at (4.5, 0) {};
					\vertex [dot, label=below:{$\kappa_{t_p}$}] (v3) at (6, -1.2) {};
					\vertex (t)  at (7, 1.5) {$t$};
					\vertex (w)  at (8.8, -0.5) {$W^- / Z / H$};
					\vertex (b)  at (8.8, -2.5) {$\bar{b} / \bar{t}$};
					\diagram* {
						(g1) -- [gluon] (v1),
						(g2) -- [gluon] (v1),
						(v1) -- [scalar, edge label=$S_8 / P_8$, edge label'=$S_1 / P_1$] (v2),
						(v2) -- [fermion] (t),
						(v2) -- [fermion, edge label=$\bar{t}_p$] (v3),
						(v3) -- [boson] (w),
						(v3) -- [anti fermion] (b)
					};
				\end{feynman}
			\end{tikzpicture}
			{(b) Gluon-Gluon Fusion}
		\end{minipage}
		\caption{Representative Feynman diagrams illustrating the dynamic topology and coupling structures for the $s$-channel production and subsequent cascade decay of a heavy resonance. \textbf{(a)} Vector states ($V_1, V_8$) produced via $q\bar{q}$ annihilation with direct renormalizable couplings ($g_q, g_V$). \textbf{(b)} Scalar and Pseudoscalar states ($S_1, P_1, S_8, P_8$) produced via loop-induced $gg$ fusion, governed by the dimension-five effective coupling ($g_{\text{eff}}/\Lambda$). In both scenarios, the vector-like top partner ($t_p$) decays democratically into a third-generation quark ($\bar{b}$ or $\bar{t}$) and an electroweak/Higgs boson ($W^-, Z$, or $H$) via standard electroweak mixing ($\kappa_{t_p}$).}
		\label{fig:feynman_split_couplings}
	\end{figure}
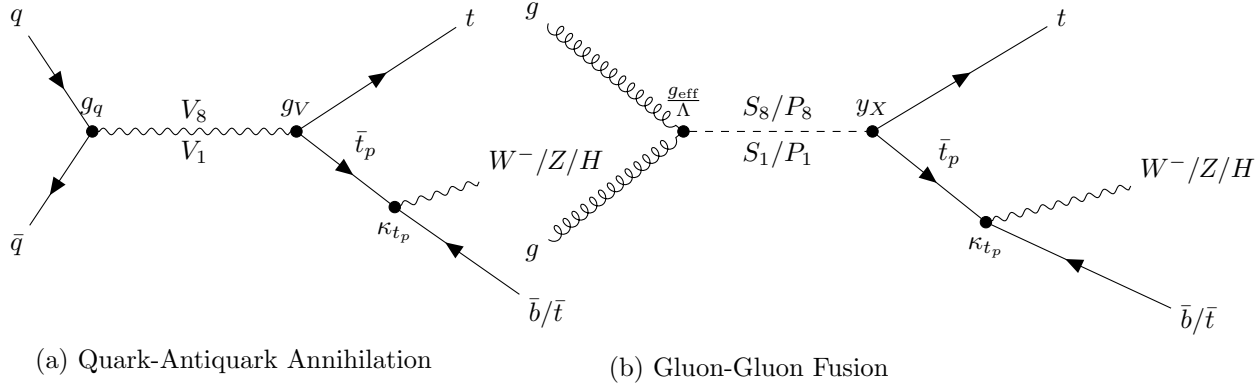

	The $s$-channel production and subsequent cascade decay topologies are illustrated in Figure~\ref{fig:feynman_split_couplings}.
	
	\section{Parton-Level Phenomenological Analysis}\label{sec3}
	
	We first analyze the signal processes strictly at the parton level to evaluate the intrinsic kinematic properties of the six spin-color resonance models. This isolates the pure hard-scattering dynamics, establishing a robust theoretical baseline prior to any subsequent phenomenological showering simulations.
	
	\subsection{Monte Carlo Generation and Benchmark Parameters}
	
	The theoretical models for the heavy resonances were implemented in \texttt{FeynRules} \cite{Alloul:2013_FeynRules} to generate Universal FeynRules Output (UFO) libraries. Leading-order (LO) Monte Carlo event generation for $pp$ collisions was performed using \texttt{MadGraph5\_aMC@NLO} v2.9.21 \cite{MadGraph, Mattelaer:2021_MadGraph}. The collisions were simulated at a center-of-mass energy of $\sqrt{s} = 13.6$~TeV, corresponding to the LHC Run 3 environment. Parton distribution functions (PDFs) were evaluated using the NNPDF sets via the LHAPDF6 framework \cite{NNPDF:2014thg, Buckley:2014ana}. The renormalization ($\mu_R$) and factorization ($\mu_F$) scales were set dynamically to the invariant mass of the heavy mediator ($\mu_R = \mu_F = M_X$).
	
	We choose a benchmark mass spectrum to probe beyond current experimental exclusions and project HL-LHC sensitivity. Setting the resonance mass to $M_X = 3$~TeV and the vector-like top partner mass to $M_{t_p} = 1.5$~TeV results in a highly boosted decay topology. To allow a direct, shape-based kinematic comparison across the different spin and color representations, the coupling constants for the $X \to t \bar{t}_p$ vertex were unified across all models. Specifically, the chiral couplings were set to $g_L = g_R = 0.2$ for the spin-1 vector states, and the Yukawa couplings were set to $y_s = y_{ps} = 0.2$ for the spin-0 states. 
	
	These specific $\mathcal{O}(0.1)$ values were deliberately chosen for two fundamental reasons. First, they ensure that the heavy resonances strictly adhere to the Narrow Width Approximation (NWA, $\Gamma_X/M_X < 5\%$), thereby avoiding finite-width distortions and preserving the perturbative unitarity of the effective interactions. Second, they yield phenomenological cross-sections that remain consistent with the non-observation limits set by current LHC searches, providing a realistic baseline for projecting HL-LHC sensitivity without artificially inflating the signal rates.
	
	\subsection{Validation of the Narrow Width Approximation and Analytical Decay Widths}
	
	To ensure the phenomenological consistency of our models, we first validate the applicability of the Narrow Width Approximation (NWA). Table~\ref{tab:width_mass_ratio} displays the dimensionless width-to-mass ratios $\Gamma_X/M_X$ across the multi-TeV mass range, demonstrating that all resonance models safely remain within the perturbative kinematic regime.
	
	\begin{table}[htbp]
		\caption{Width-to-mass ratios $\Gamma_X/M_X$ (in \%) corresponding to the heavy resonance states, demonstrating strict adherence to the Narrow Width Approximation ($\Gamma/M < 5\%$).}
		\label{tab:width_mass_ratio}
		\centering
		\small 
		\renewcommand{\arraystretch}{1.2} 
		\begin{tabular}{ccccccc}
			\toprule
			\textbf{Mass [GeV]} & \textbf{$V_8$ (\%)} & \textbf{$S_8$ (\%)} & \textbf{$P_8$ (\%)} & \textbf{$V_1$ (\%)} & \textbf{$S_1$ (\%)} & \textbf{$P_1$ (\%)} \\ 
			\midrule
			\textbf{2000} & 0.75\% & 0.05\% & 0.07\% & 4.49\% & 0.28\% & 0.38\% \\
			\textbf{2500} & 0.77\% & 0.11\% & 0.12\% & 4.65\% & 0.58\% & 0.68\% \\
			\textbf{3000} & 0.79\% & 0.16\% & 0.17\% & 4.74\% & 0.85\% & 0.93\% \\
			\textbf{3500} & 0.80\% & 0.20\% & 0.21\% & 4.79\% & 1.09\% & 1.15\% \\
			\textbf{4000} & 0.80\% & 0.25\% & 0.25\% & 4.83\% & 1.31\% & 1.36\% \\
			\textbf{4500} & 0.81\% & 0.29\% & 0.30\% & 4.85\% & 1.54\% & 1.58\% \\
			\textbf{5000} & 0.81\% & 0.34\% & 0.34\% & 4.86\% & 1.77\% & 1.80\% \\
			\textbf{5500} & 0.81\% & 0.38\% & 0.39\% & 4.88\% & 2.00\% & 2.03\% \\
			\bottomrule
		\end{tabular}
	\end{table}
	
	The $\Gamma_X/M_X$ ratio remains small across the investigated mass range. The maximum ratio occurs in the color-singlet vector configuration ($V_1$), reaching approximately $4.88\%$ for a $5.5$~TeV mass shell, due to its unsuppressed chiral couplings to light quarks. The color-octet vector ($V_8$) and all spin-0 scalar/pseudoscalar models maintain ratios below $2.5\%$, suppressed by their specific color and loop-induced structures. 
	
	Since $\Gamma_X/M_X < 5\%$ for all models, the NWA is justified \cite{Berdine:2007_NWA}. In this kinematic regime, the $2 \to 2$ scattering amplitude factorizes into an on-shell production cross-section multiplied by the decay branching ratio. For the color-singlet vector channel, the partonic cross-section is:
	\begin{equation}
		\hat{\sigma}(q\bar{q} \to t\bar{t}_p) \approx \hat{\sigma}(q\bar{q} \to V_1) \times \mathcal{B}(V_1 \to t\bar{t}_p)
	\end{equation}
	where $\mathcal{B}$ denotes the branching ratio of the vector resonance decaying into the top quark and top partner final state. This factorization simplifies the calculation by allowing us to safely neglect off-shell interference effects \cite{Uhlemann:2008_NWA}.
	
	To evaluate the decay dynamics driving this branching ratio, we analytically compute the two-body partial decay width of the vector resonance $V_1$ into a top quark and a top partner:
	\begin{equation}
		\Gamma(V_1 \to t\bar{t}_p) = \frac{\lambda^{1/2}(M_{V_1}^2, m_t^2, m_{t_p}^2)}{16 \pi M_{V_1}^3} \overline{|\mathcal{M}|^2} \,,
		\label{eq:decay_width}
	\end{equation}
	where $M_{V_1}$, $m_t$, and $m_{t_p}$ denote the masses of the vector resonance, the SM top quark, and the top partner, respectively. The kinematic K\"all\'en function \cite{Byckling:1973, PDG:2022} is explicitly defined as $\lambda(x, y, z) = x^2 + y^2 + z^2 - 2xy - 2xz - 2yz$.
	
	The squared invariant matrix element $\overline{|\mathcal{M}|^2}$, averaged over initial spins and colors and summed over final states, is derived as:
	\begin{align}
		\overline{|\mathcal{M}|^2} &= \frac{1}{M_{V_1}^2} \bigg[ (g_{1L}^2 + g_{1R}^2) \Big( (m_t^2 - m_{t_p}^2)^2 + M_{V_1}^2 (m_t^2 + m_{t_p}^2) - 2M_{V_1}^4 \Big) \nonumber \\
		&\quad - 12 g_{1L} g_{1R} m_t m_{t_p} M_{V_1}^2 \bigg] \,.
		\label{eq:squared_amplitude}
	\end{align}
	This structure is in full agreement with standard quantum field theory derivations for massive vector decays \cite{Peskin:1995_QFT, Buchkremer:2013}. The interference term proportional to $g_{1L} g_{1R}$ requires fermion masses to induce chirality flipping, consistent with expectations for interactions not involving massless final states.
	
	\vspace{0.5cm}
	\noindent\textbf{Negligibility of Signal-Background Interference:} \\
	In our Monte Carlo simulations, the interference between the $s$-channel BSM signal and the SM background continuum ($gg/q\bar{q} \to t\bar{t}$ and $t\bar{t} + \text{jets}$) is neglected ($|\mathcal{M}_{\text{tot}}|^2 \approx |\mathcal{M}_{\text{sig}}|^2 + |\mathcal{M}_{\text{bkg}}|^2$). This is justified for two reasons. First, as shown in Table~\ref{tab:width_mass_ratio}, the resonances operate in the narrow-width regime ($\Gamma_X/M_X < 5\%$), making the kinematically integrated interference term marginal over the broad SM continuum \cite{Carena:2016_Interference, Berdine:2007_NWA}. Second, fundamental color mismatches—such as color-singlet states ($\mathbf{1}$) interfering with the color-octet QCD background ($\mathbf{8}$)—cause the color-traced interference terms ($\text{Tr}(T^a) = 0$) to vanish at leading order \cite{Baur:2001_Zprime}.
	
	\subsection{Exact Analytical Matrix Elements and Partonic Cross-Sections}
	
	To understand the angular correlations responsible for spin discrimination, we derive the fully contracted squared matrix elements $\overline{|\mathcal{M}|^2}$ and the partonic cross-sections $\hat{\sigma}(s)$ for the $2 \to 2$ scattering processes \cite{Gao:2010_Spin, Barger:1987_ColliderPhysics}.
	
	We express the differential cross-section with respect to the scattering angle $\theta$ in the center-of-mass frame. For the spin-1 vector mediators ($V_1, V_8$), helicity conservation along the vector propagator leads to an angular dependence:
	\begin{equation}
		\frac{d\hat{\sigma}_V}{d\cos\theta} \propto \mathcal{F}_{\text{kin}}(s) + \mathcal{G}_{\text{kin}}(s) {\cos^2\theta}
	\end{equation}
	This $\cos^2\theta$ structure causes the decay products to be emitted preferentially at specific angles, smearing the kinematic edges when boosted to the transverse plane.
	
	For the spin-0 mediators—both CP-even scalar ($S$) and CP-odd pseudoscalar ($P$)—the interactions are purely $s$-dependent, and the differential cross-section has no angular dependence:
	\begin{equation}
		\frac{d\hat{\sigma}_{S,P}}{d\cos\theta} \propto \mathcal{K}_{\text{kin}}(s) \,.
	\end{equation}
	The $s$-dependent kinematic coefficients for the vector mediator are:
	\begin{align}
		\mathcal{F}_{\text{kin}}(s) &= g_{1L} g_{1R} m_t m_{t_p} s + \frac{g_{1L}^2 + g_{1R}^2}{8} \Big[ s^2 - (m_t^2 - m_{t_p}^2)^2 \Big] \,, \\
		\mathcal{G}_{\text{kin}}(s) &= \frac{g_{1L}^2 + g_{1R}^2}{8} \lambda(s, m_t^2, m_{t_p}^2) \,.
	\end{align}
	The angular coefficient $\mathcal{G}_{\text{kin}}(s)$ is proportional to the K\"all\'en function $\lambda$, which drives the momentum broadening. 
	
	For the spin-0 mediators, the isotropic threshold functions $\mathcal{K}_{\text{kin}}(s)$ are:
	\begin{align}
		\mathcal{K}_{\text{kin}}^S(s) &= s^2 \Big( s - (m_t + m_{t_p})^2 \Big) \quad \text{for CP-even } (S) \,, \\
		\mathcal{K}_{\text{kin}}^P(s) &= s^2 \Big( s - (m_t - m_{t_p})^2 \Big) \quad \text{for CP-odd } (P) \,.
	\end{align}
	This implies that scalar and pseudoscalar resonances decay isotropically in their rest frames. Boosting this isotropic distribution to the transverse plane produces a sharp Jacobian peak near the kinematic endpoint $p_T^{\text{max}} \approx \lambda^{1/2}(M_{X}^2, m_t^2, m_{t_p}^2) / 2M_{X}$ \cite{Byckling:1973}. Therefore, the angular terms in the matrix elements determine the $p_T$ shape, establishing it as an observable for spin discrimination.
	
	The inclusive hadronic cross-section for $pp \to t\bar{t}_p$ is obtained by convoluting the parton-level cross-sections with the Parton Distribution Functions (PDFs). Based on the QCD factorization theorem \cite{Collins:1989_Factorization, Campbell:2006_Primer}, the hadronic cross-section is:
	\begin{equation}
		\sigma(pp \to t\bar{t}_p) = \sum_{i,j} \int_{\tau_{\text{min}}}^{1} dx_i \int_{\tau_{\text{min}}/x_i}^{1} dx_j \, f_{i/p}(x_i, \mu_F^2) \, f_{j/p}(x_j, \mu_F^2) \, \hat{\sigma}_{ij \to t\bar{t}_p}(\hat{s}, \mu_R^2, \mu_F^2) \,,
		\label{eq:hadronic_cross_section}
	\end{equation}
	where $i, j \in \{g, q, \bar{q}\}$ denote the initial interacting partons, $x_i$ and $x_j$ represent the energy fractions (longitudinal momentum fractions) of the parent protons carried by these partons, and $f_{i/p}$ are the PDFs \cite{CTEQ:1995_Handbook}. Here, $\hat{\sigma}_{ij \to t\bar{t}_p}$ represents the partonic cross-section calculated perturbatively for the intermediate state, without explicitly including the resonance in the final state definition. The integration threshold is $\tau_{\text{min}} = (m_t + m_{t_p})^2 / s$. This formulation accommodates both $q\bar{q}$ annihilation (dominant for $V_1$) and gluon-gluon fusion (dominant for $S, P$, and $V_8$).
	
	\subsection{Production Cross-Sections and PDF Luminosity Dynamics}
	\label{subsec:prod_cross}
	
\begin{figure}[ht]
		\centering
		\subfigure{
			\includegraphics[width=0.30\textwidth]{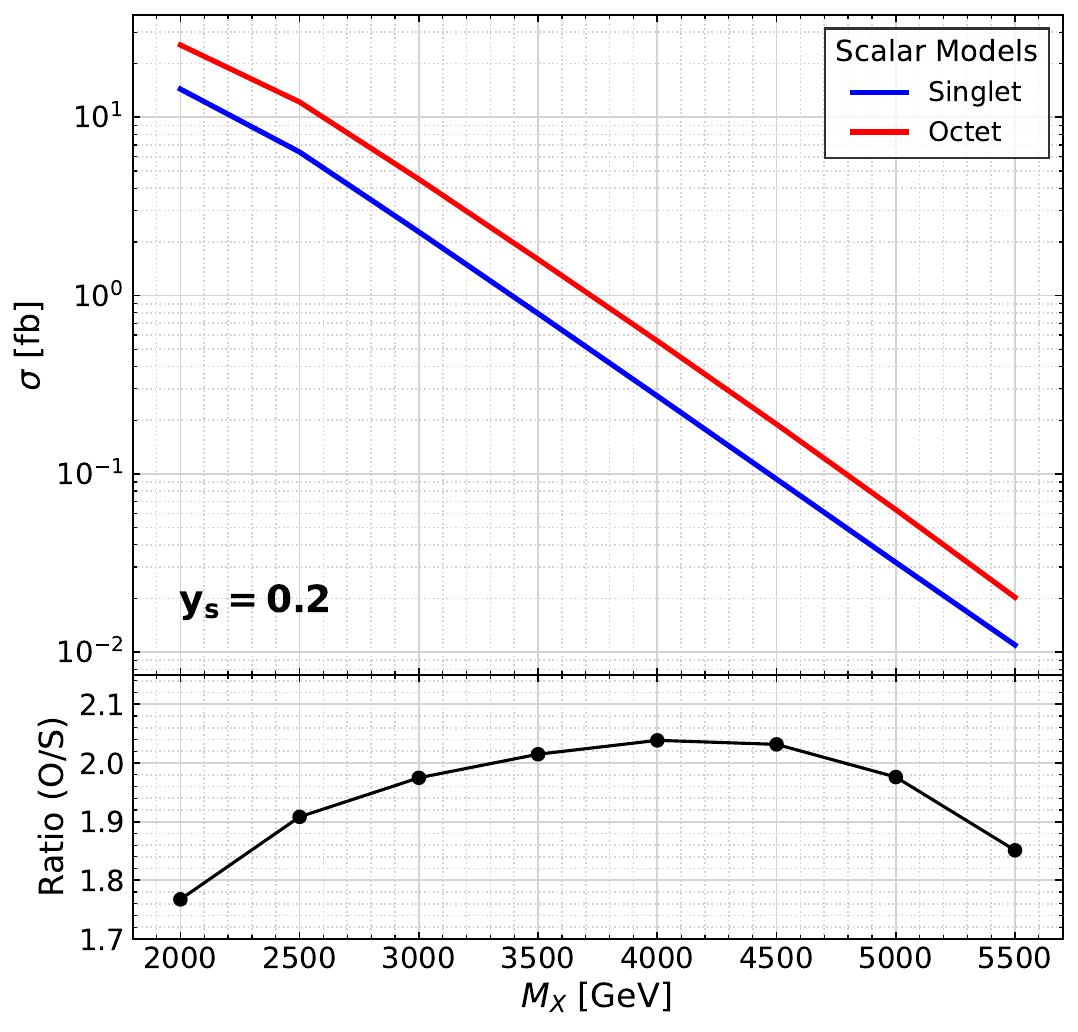}
			\label{fig:cross_scalar}
		}
		\hfill
		\subfigure{
			\includegraphics[width=0.30\textwidth]{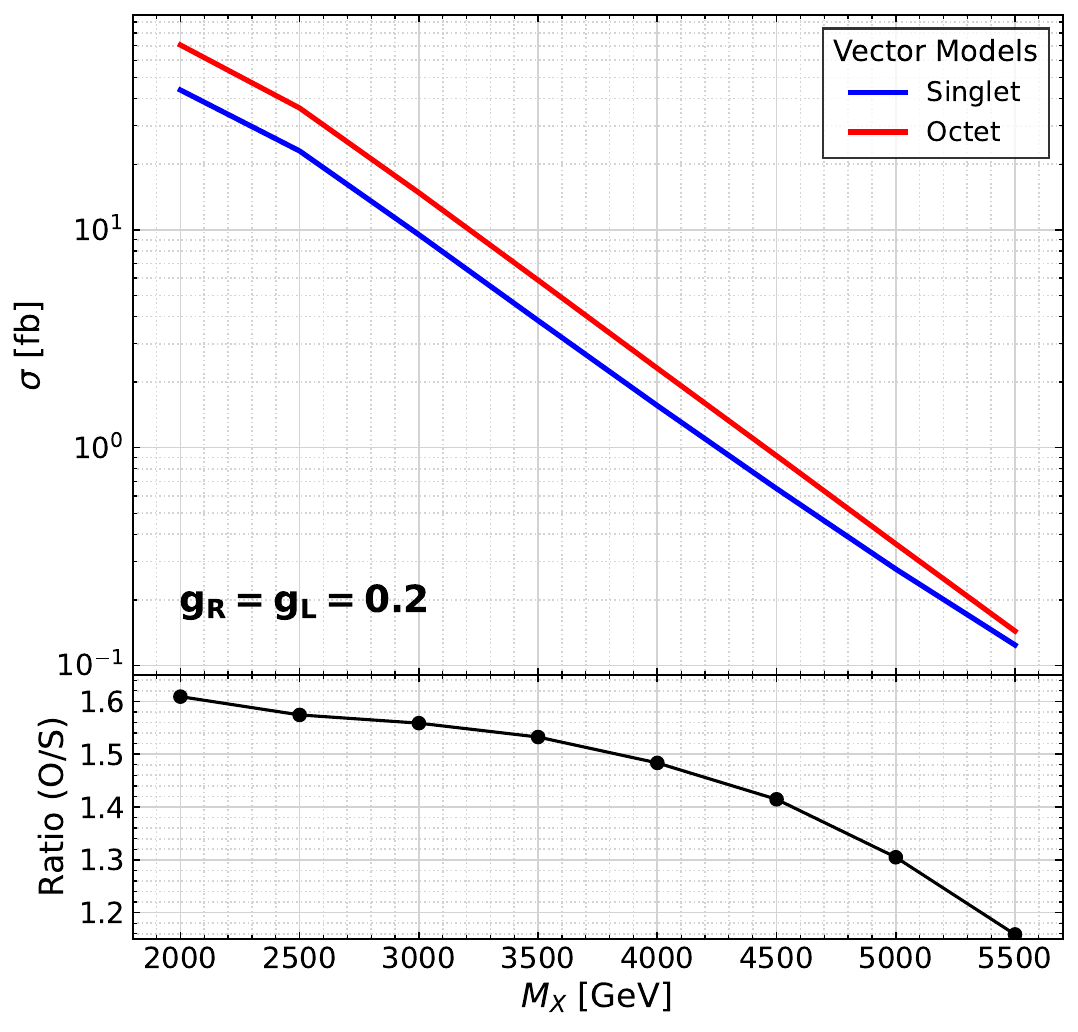}
			\label{fig:cross_vector}
		}
		\hfill
		\subfigure{
			\includegraphics[width=0.30\textwidth]{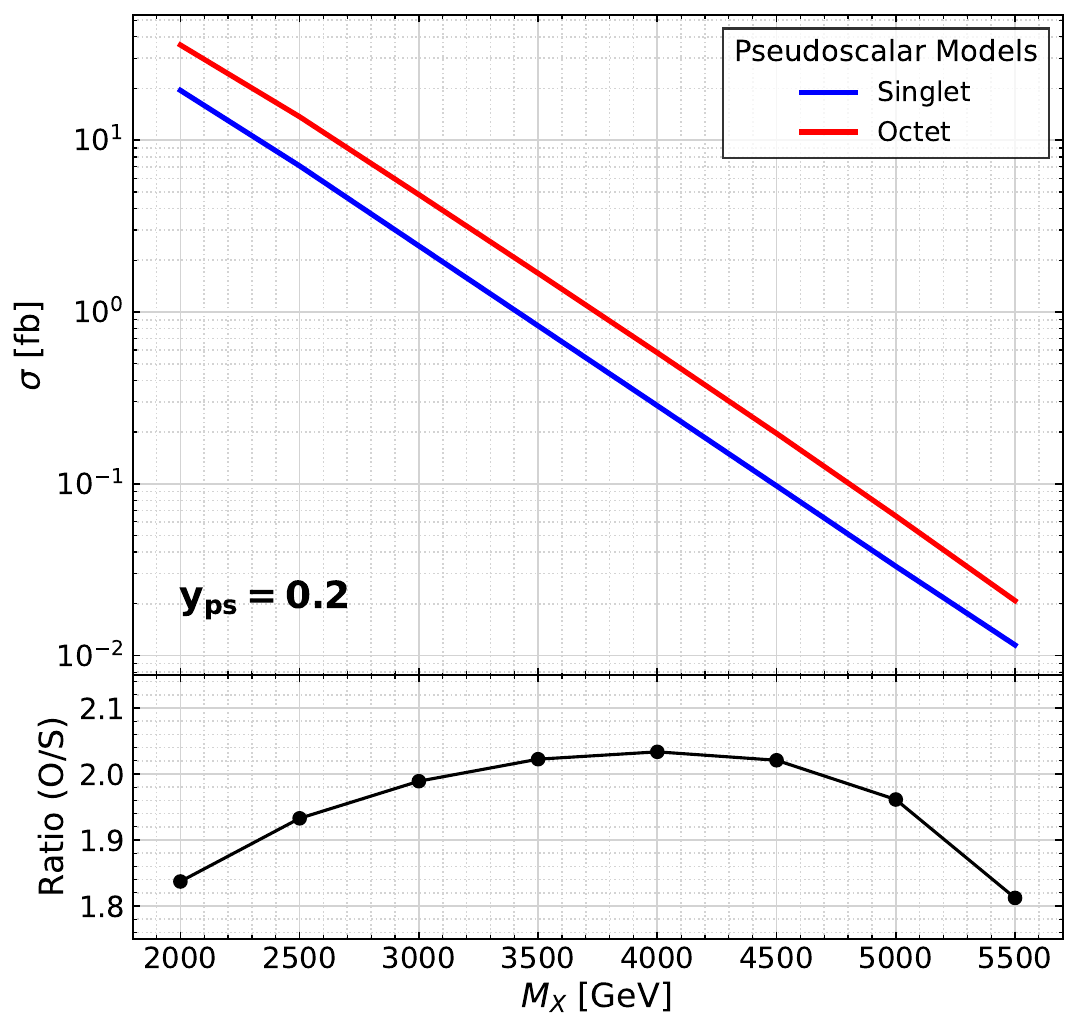}
			\label{fig:cross_pseudo}
		}
		\caption{Cross-sections $\sigma$ (in fb) as a function of the resonance mass $M_X$ for the Scalar (left), Vector (center), and Pseudo-scalar (right) models. Each panel compares the color-singlet and color-octet representations. The bottom sub-panels display the cross-section ratio (Octet/Singlet), highlighting the dynamic enhancement driven by the QCD color factors.}
		\label{fig:cross_sections}
	\end{figure}

	We summarize the inclusive production cross-sections ($\sigma$) for the heavy mediators in Table~\ref{tab:cross_sections} and Figure~\ref{fig:cross_sections}.
	
	\begin{table}[htbp]
		\caption{Cross-sections (in fb) for the six heavy resonance models at $\sqrt{s} = 13.6$~TeV as a function of the mediator mass.}
		\label{tab:cross_sections}
		\centering
		\small
		\renewcommand{\arraystretch}{1.2}
		\begin{tabular}{ccccccc}
			\toprule
			\textbf{Mass [GeV]} & \textbf{$V_8$ (fb)} & \textbf{$S_8$ (fb)} & \textbf{$P_8$ (fb)} & \textbf{$V_1$ (fb)} & \textbf{$S_1$ (fb)} & \textbf{$P_1$ (fb)} \\ 
			\midrule
			\textbf{2000} & 70.5070 & 25.4500 & 35.8540 & 43.8190 & 14.3985 & 19.5165 \\
			\textbf{2500} & 36.2050 & 12.1819 & 13.7013 & 22.9960 & 6.3838  & 7.0887  \\
			\textbf{3000} & 14.8027 & 4.4950  & 4.8249  & 9.4963  & 2.2762  & 2.4257  \\
			\textbf{3500} & 5.8689  & 1.5964  & 1.6810  & 3.8296  & 0.7923  & 0.8312  \\
			\textbf{4000} & 2.3123  & 0.5566  & 0.5792  & 1.5587  & 0.2730  & 0.2848  \\
			\textbf{4500} & 0.9172  & 0.1897  & 0.1964  & 0.6484  & 0.0934  & 0.0972  \\
			\textbf{5000} & 0.3614  & 0.0629  & 0.0649  & 0.2770  & 0.0319  & 0.0331  \\
			\textbf{5500} & 0.1438  & 0.0203  & 0.0209  & 0.1241  & 0.0110  & 0.0115  \\
			\bottomrule
		\end{tabular}
	\end{table}
	
	The color-octet states ($V_8, S_8, P_8$) yield significantly larger cross-sections than their color-singlet counterparts ($V_1, S_1, P_1$). This dynamic enhancement is governed by two fundamental mechanisms. First, the $SU(3)_C$ color trace algebra introduces larger color factors directly into the squared matrix elements of the octet states. Second, and more crucially at $\sqrt{s} = 13.6$~TeV, the gluon-gluon parton luminosity overwhelmingly dominates over the quark-antiquark ($q\bar{q}$) density at multi-TeV energy scales \cite{Campbell:2006_Primer}. Color-octet states couple directly to initial-state gluons via unsuppressed three-point vertices, whereas color-singlet vectors ($V_1$) are produced primarily via $q\bar{q}$ annihilation, leading to lower overall rates despite their dimension-four couplings \cite{Simmons:2013_Coloron}.
	
	These cross-sections are evaluated at Leading Order (LO) in perturbative QCD. It is well established in the literature that including Next-to-Leading Order (NLO) QCD corrections introduces a positive $K$-factor ($\sigma_{\text{NLO}} = K \cdot \sigma_{\text{LO}}$). For massive color-octet resonances produced via gluon-gluon fusion, theoretical NLO calculations yield a $K$-factor typically ranging from 1.2 to 1.3 \cite{Zhu:2012_NLO, Freitas:2017_ColorOctetNLO}. This implies that NLO effects enhance the inclusive production rate by 20\% to 30\% while simultaneously reducing the unphysical dependence on the renormalization and factorization scales. As such, our LO cross-sections provide a safe lower bound for the expected signal yields. If a signal is viable at LO, the inclusion of higher-order corrections will inherently improve the projected HL-LHC sensitivity.
	
	Therefore, rather than performing computationally prohibitive full NLO Monte Carlo generation for these multi-TeV resonances, adopting established NLO K-factors from the literature provides a highly reliable and conservative baseline to evaluate the kinematic improvements introduced by our deep learning framework.
	
	\subsection{Differential Kinematics and Spin Discrimination}
	
	To examine these kinematic properties, we analyze the differential distributions of the final-state particles. All distributions are normalized to unity to compare shape differences directly.
	
	\begin{figure*}[htbp]
		\centering
		\includegraphics[width=0.95\textwidth]{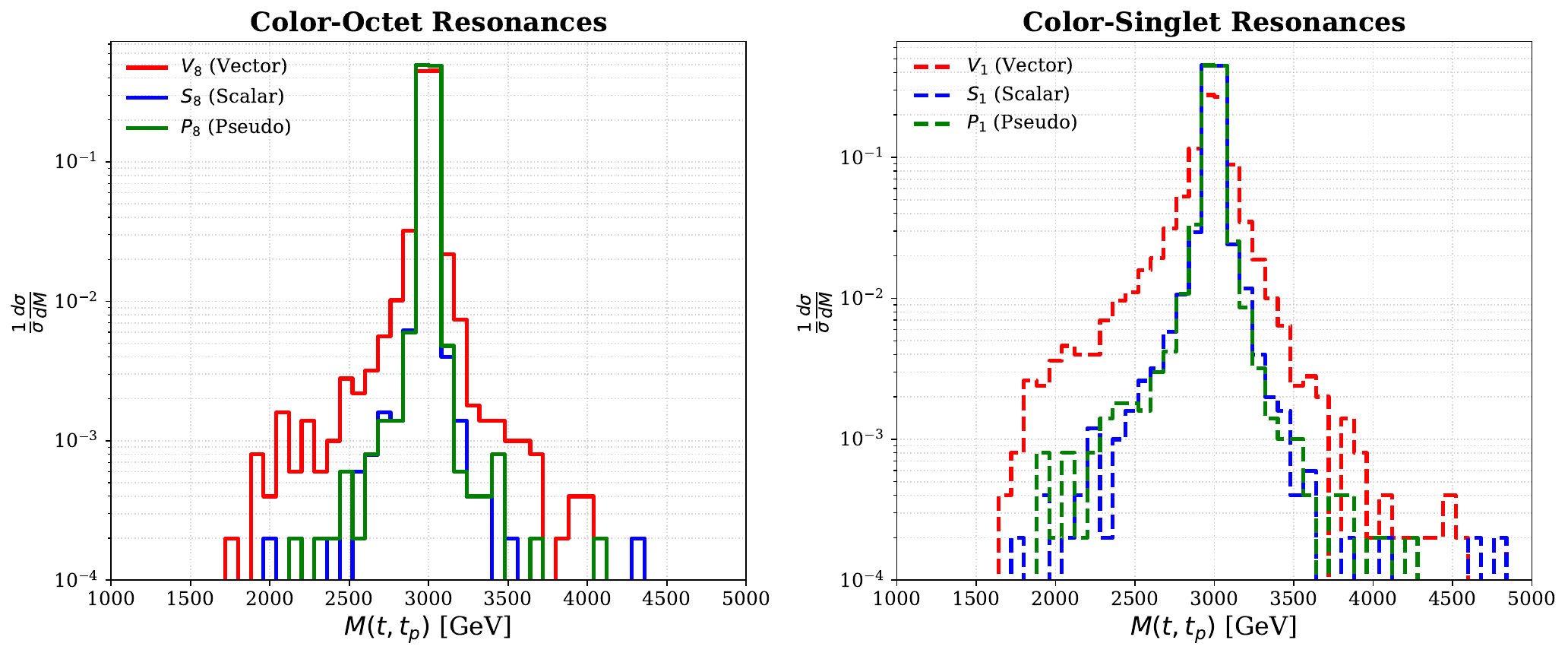}
		\caption{Parton-level normalized invariant mass distributions $M(t, t_p)$ for the six BSM scenarios at $\sqrt{s} = 13.6$~TeV. All benchmarks demonstrate a sharp resonance peak centered at 3~TeV.}
		\label{fig:mtt_dist}
	\end{figure*}
	
	Figure~\ref{fig:mtt_dist} illustrates the invariant mass distribution of the top-sector system, $M(t, t_p)$. While this variable serves as the primary discovery channel by displaying a sharp Breit-Wigner peak at $M_X = 3$~TeV, it exhibits complete kinematic degeneracy regarding the mediator's spin and color representations. Breaking this degeneracy requires analyzing the normalized differential cross-section with respect to the transverse momentum ($p_T$) and pseudorapidity ($\eta$) profiles.
	
	\begin{figure*}[htbp]
		\centering
		\includegraphics[width=0.95\textwidth]{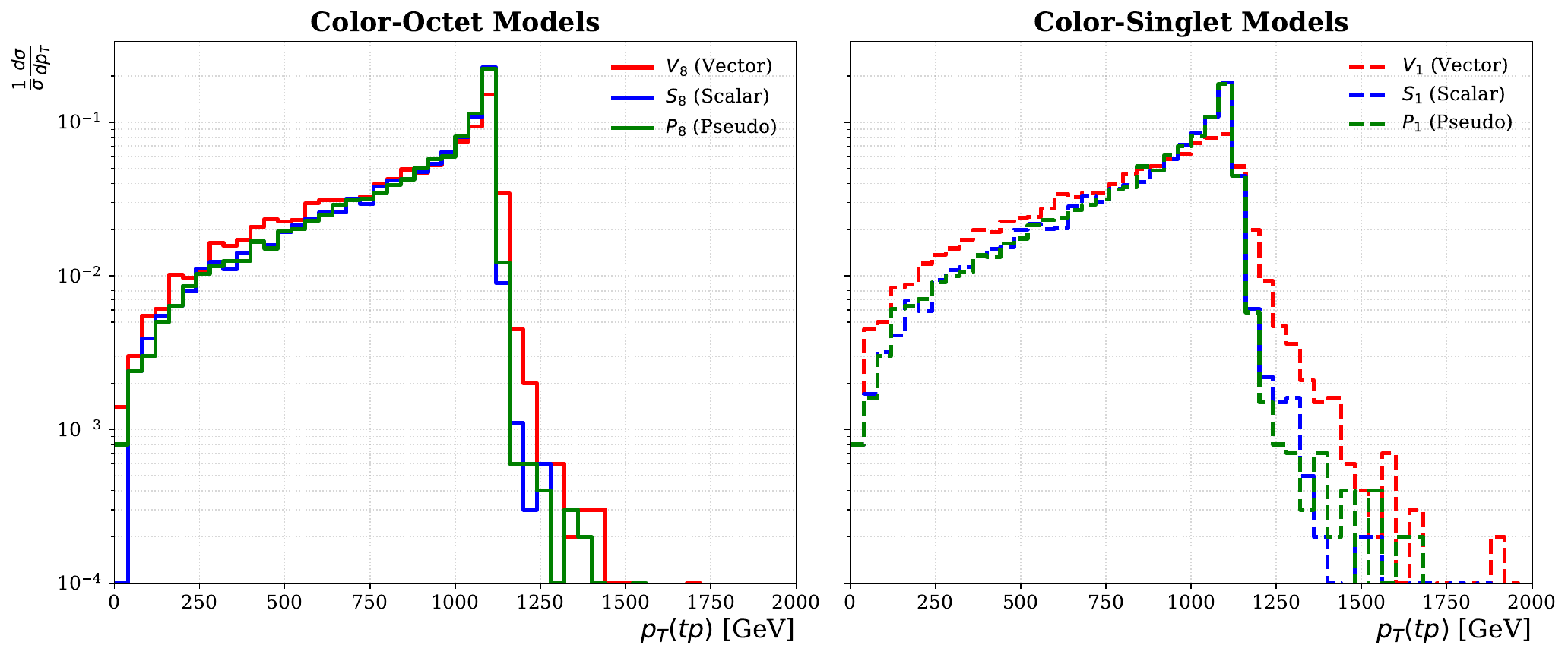}
		\caption{Parton-level normalized differential cross-sections as a function of transverse momentum, $\frac{1}{\sigma}\frac{d\sigma}{dp_T}$, for the vector-like top partner ($t_p$).}
		\label{fig:pt_comparison}
	\end{figure*}
	
	\begin{figure*}[htbp]
		\centering
		\includegraphics[width=0.95\textwidth]{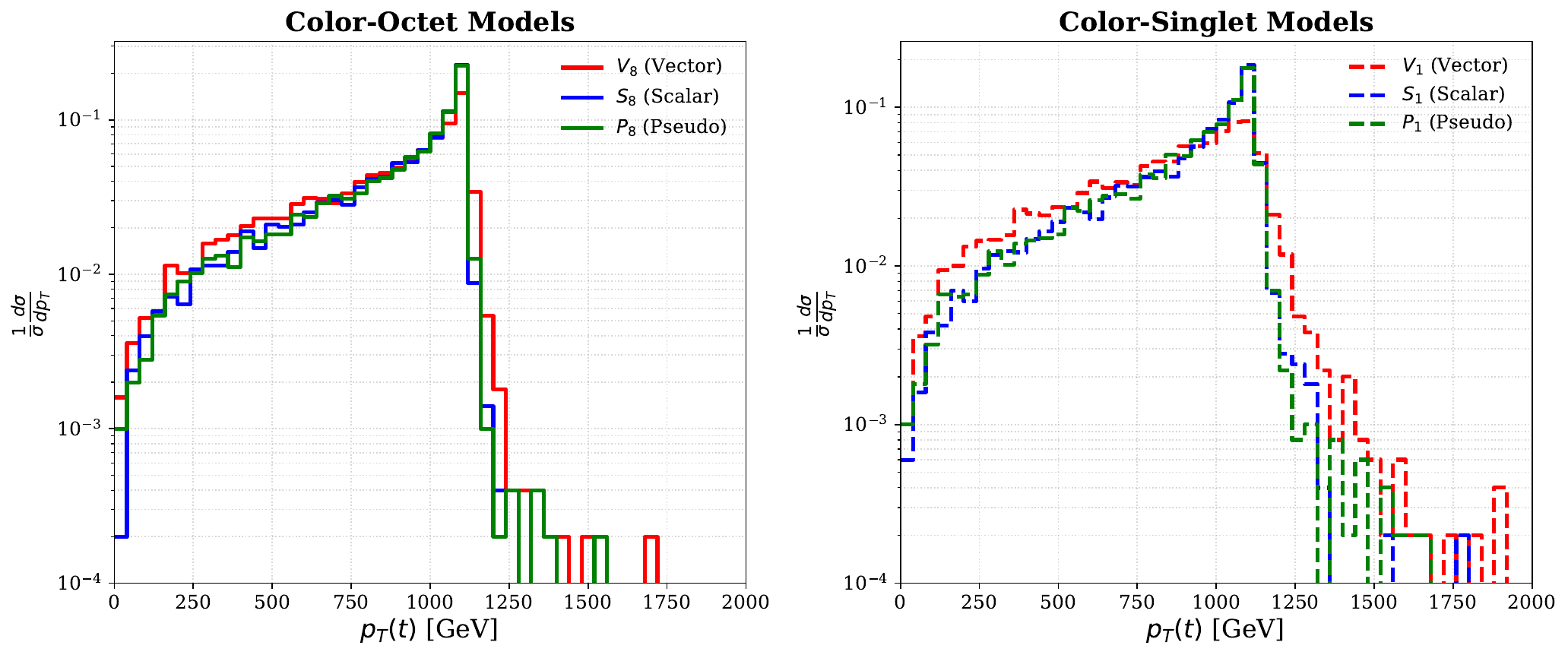}
		\caption{Parton-level normalized differential cross-sections as a function of transverse momentum, $\frac{1}{\sigma}\frac{d\sigma}{dp_T}$, for the SM top quark ($t$).}
		\label{fig:top_pt_comparison}
	\end{figure*}
	
	The normalized differential cross-sections as a function of transverse momentum, $\frac{1}{\sigma}\frac{d\sigma}{dp_T}$, for the top partner $p_T(t_p)$ and the top quark $p_T(t)$ (Figures~\ref{fig:pt_comparison} and \ref{fig:top_pt_comparison}) establish a robust framework for spin discrimination. The physical origin of this discrimination lies in the analytical structure of the matrix elements derived in Equations 3.6-3.9. Spin-0 mediators ($S, P$) decay isotropically in their rest frames ($\frac{d\hat{\sigma}}{d\cos\theta} \propto \text{const}$). When boosted to the transverse plane, the kinematic change of variables from the scattering angle to transverse momentum introduces a singularity factor. This mathematical transformation translates the flat angular isotropy into a sharp, highly localized Jacobian peak near the kinematic threshold $p_T^{\text{max}} \approx \lambda^{1/2}(M_{X}^2, m_t^2, m_{t_p}^2) / 2M_{X}$. Conversely, for spin-1 mediators, helicity conservation strictly imposes a $\cos^2\theta$ angular dependence. This forces the decay products to be emitted preferentially along the beam axis (forward/backward regions where $\theta \to 0, \pi$) and heavily suppresses emissions at central transverse angles ($\theta \approx \pi/2$) where $p_T$ is naturally maximized. As a consequence, the Jacobian peak is completely washed out, heavily smearing the transverse momentum distribution and resulting in a significantly broader profile.
	
	\begin{figure*}[htbp]
		\centering
		\includegraphics[width=0.95\textwidth]{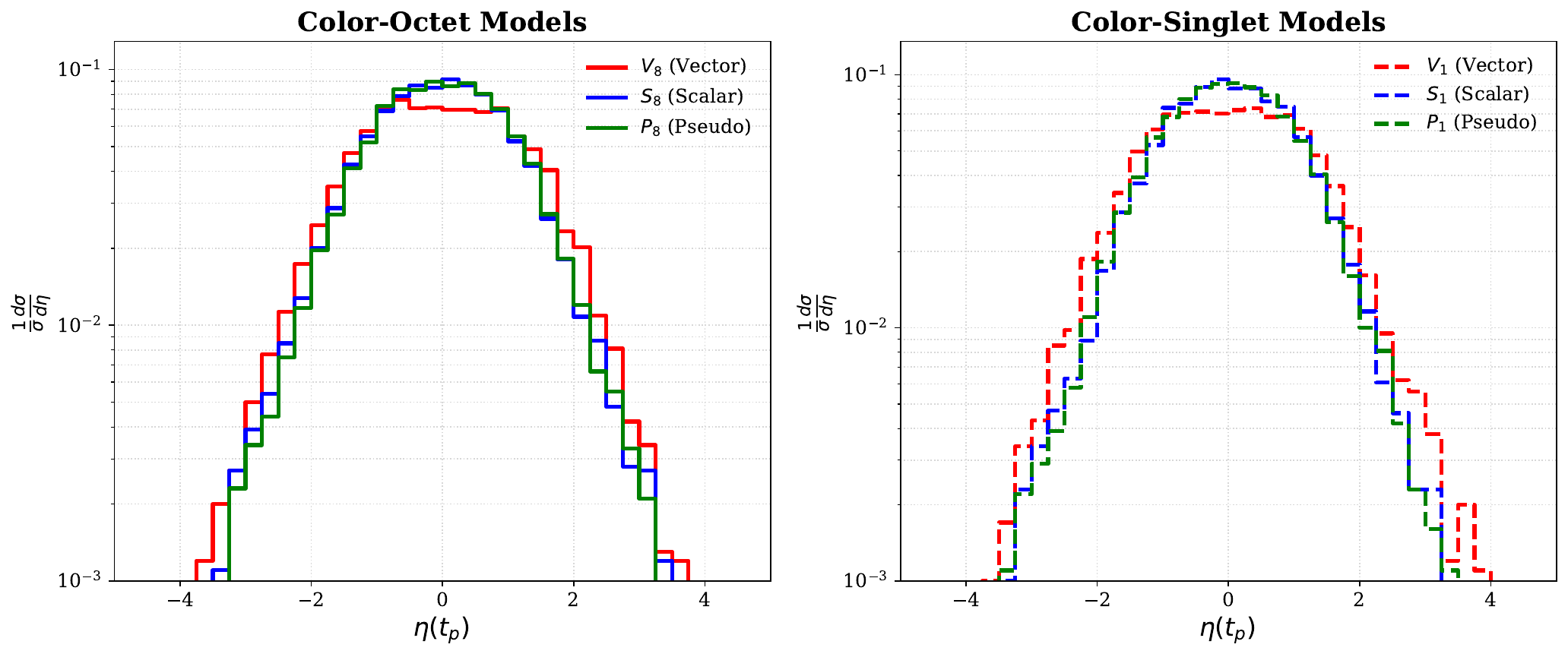}
		\caption{Parton-level normalized pseudorapidity ($\eta$) distributions of the vector-like top partner ($t_p$).}
		\label{fig:tp_eta_comparison}
	\end{figure*}
	
	Furthermore, the pseudorapidity profiles of the top partner, $\eta(t_p)$ (Figure~\ref{fig:tp_eta_comparison}), demonstrate a symmetric topology strictly confined within the central detector barrel ($|\eta| < 2.5$). Producing a massive 3~TeV state at $\sqrt{s} = 13.6$~TeV demands initial partons with high and highly symmetric momentum fractions ($x_1 \approx x_2 \approx M_X/\sqrt{s} \approx 0.22$). This naturally leads to a vanishing longitudinal boost ($x_F \approx 0$), forcing the highly energetic decay products to be emitted centrally and at wide angles relative to the beam axis, an optimal topology for central tracking detectors.
	
	\section{Detector-Level Simulation and Event Selection}\label{sec4}
	
	After examining the parton-level kinematics, we transition to a realistic experimental framework. In the hadronic environment of the Large Hadron Collider (LHC), the final-state particles undergo QCD bremsstrahlung, showering, and hadronization before interacting with the detector. Simulating these non-perturbative effects and finite calorimetric resolutions is necessary to evaluate the expected discovery reach. To maximize signal acceptance for multi-TeV cascades, this analysis focuses strictly on the inclusive fully-hadronic decay channels of the top partner to maximize kinematic reconstruction.
	
	\subsection{Hadronization and Fast-Detector Reconstruction}
	
	To transition from the ideal parton-level kinematics to a realistic experimental framework, we simulate the non-perturbative effects inherent in the hadronic environment of the Large Hadron Collider (LHC). The parton-level Les Houches Events (LHE) were processed with \texttt{Pythia 8.2} \cite{Pythia8} to account for initial- and final-state radiation (ISR/FSR), multiple parton interactions (MPI), and hadronization. The resulting stable-particle final states were passed through \texttt{Delphes 3} \cite{Delphes} for fast-detector simulation, employing standard ATLAS/CMS parameterizations for tracking and calorimeter energy smearing. 
	
	Given the massive energy scale of the signal ($M_X = 3$~TeV), the decay products are highly collimated. To reconstruct these boosted topologies, we utilize the anti-$k_t$ algorithm via \texttt{FastJet} \cite{FastJet} with a cone radius of $R = 0.8$ to capture the full hadronic cascade in "fat-jets" ($j_1, j_2$). We assume that the degradation in jet mass resolution typically expected from pile-up is mitigated by grooming techniques such as Soft Drop \cite{Larkoski:2014_SoftDrop}, restoring the substructure kinematics to the baseline environment.
	
	\subsection{Event Topology and Selection Observables}
	
	Before applying the selection strategy, we characterize the global kinematic event topology. The signal models are driven by the hard-scattering scale of a 3~TeV resonance, while the Standard Model $t\bar{t}$ background follows a rapidly falling spectrum. We utilize $H_T$, $\slashed{E}_T$, and $p_T(j_1)$ as the primary observables to resolve this spectral separation.
	
	\begin{figure}[htbp]
		\centering
		\includegraphics[width=0.95\textwidth]{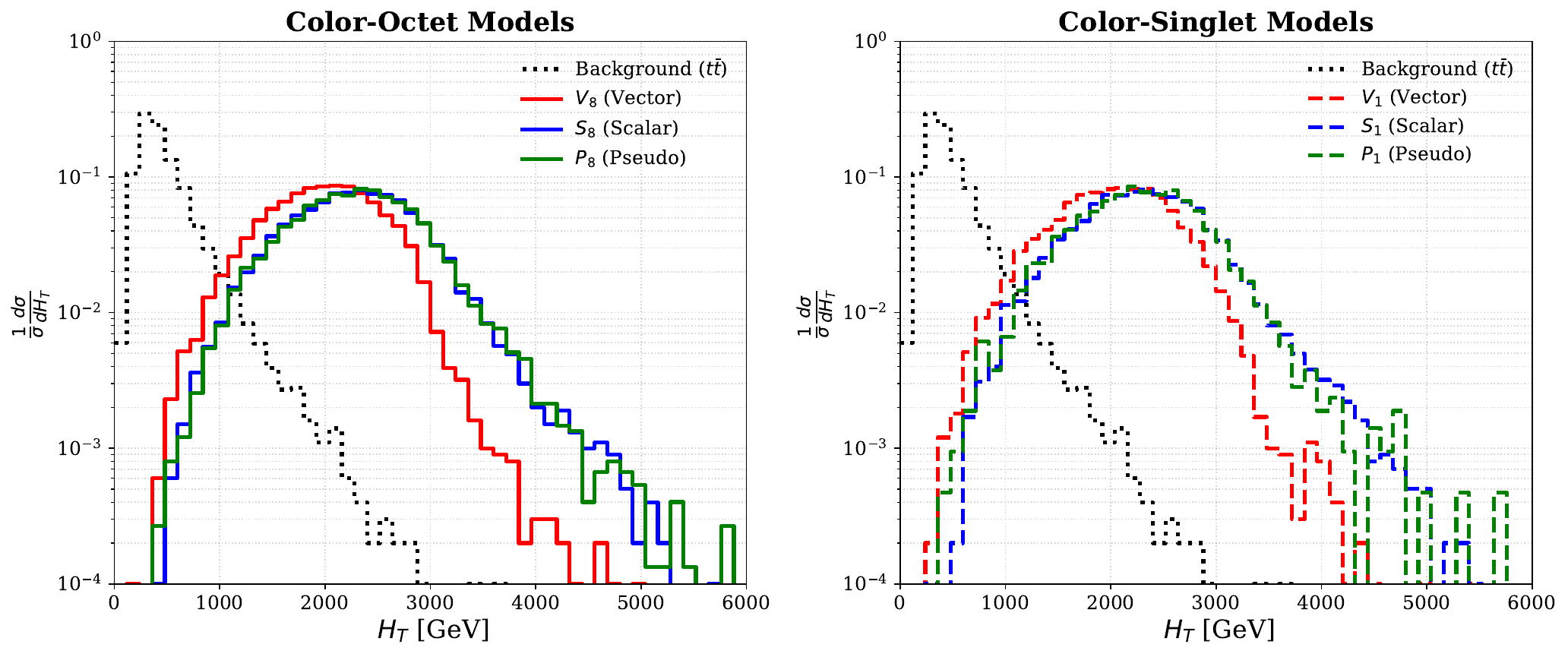}
		\caption{Normalized distributions of the scalar sum of transverse momenta, $H_T$. The high-mass resonance transfers massive energy to the final state, shifting the $H_T$ distribution significantly beyond 2~TeV, away from the SM continuum.}
		\label{fig:ht_distribution}
	\end{figure}
	
	The Hadronic Scalar Sum of transverse momenta ($H_T$) acts as a robust proxy for the hard-scattering scale. As shown in Figure~\ref{fig:ht_distribution}, the massive resonance yields a broad signal distribution peaking above 2~TeV. In contrast, the SM $t\bar{t}$ continuum is heavily suppressed by the parton luminosity fall-off at high Bjorken-$x$, reinforcing $H_T$ as a primary discriminant.
	
	\begin{figure}[htbp]
		\centering
		\includegraphics[width=0.95\textwidth]{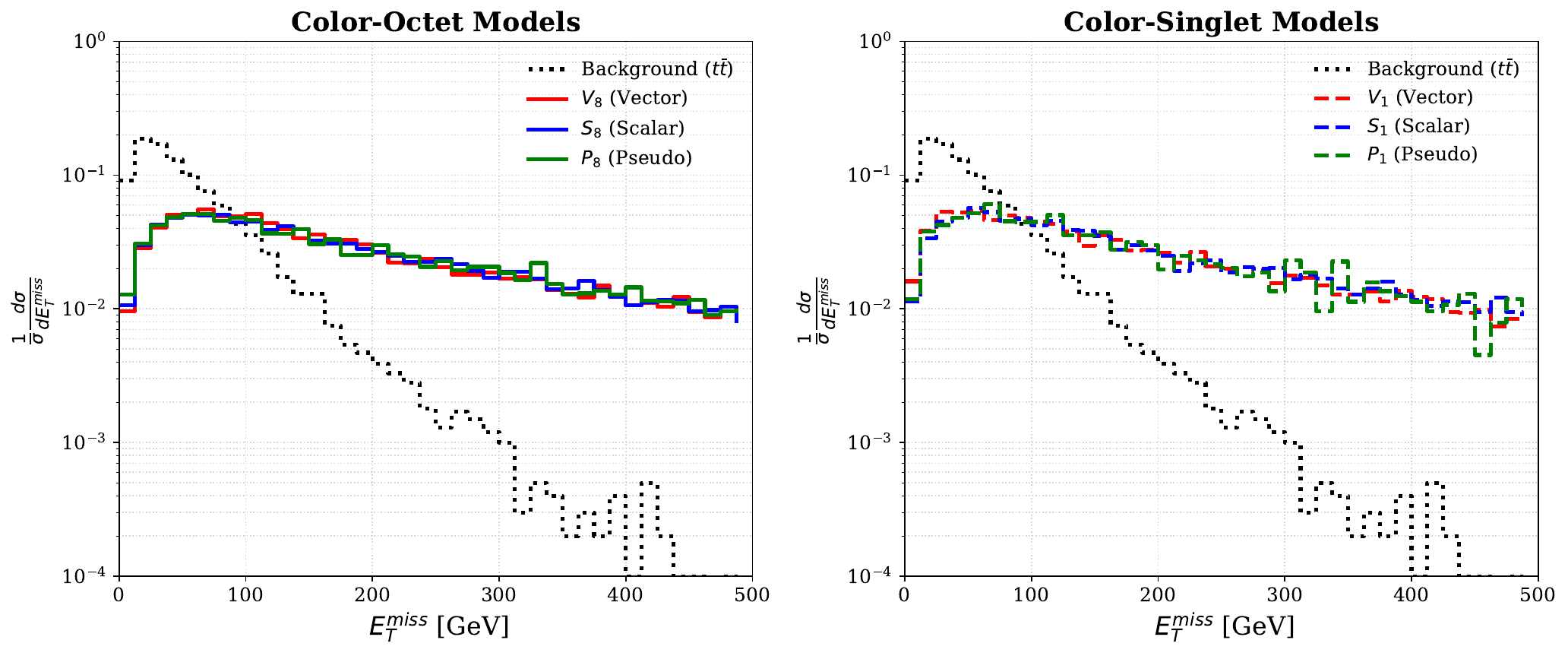}
	 
		\caption{Normalized distributions of the Missing Transverse Energy ($\slashed{E}_T$). The signal exhibits a hard tail at high $\slashed{E}_T$ due to the high Lorentz boost imparted to neutrinos from electroweak sub-decays.}
		\label{fig:met_distribution}
	\end{figure}
	
	Similarly, the Missing Transverse Energy ($\slashed{E}_T$) in Figure~\ref{fig:met_distribution} highlights the impact of the 3~TeV Lorentz boost. Even in predominantly hadronic final states, neutrinos from top/W sub-decays inherit significant transverse momentum, manifesting as a high-energy tail in the signal spectrum that contrasts with the softer SM background.
	
	\begin{figure}[htbp]
		\centering
		\includegraphics[width=0.95\textwidth]{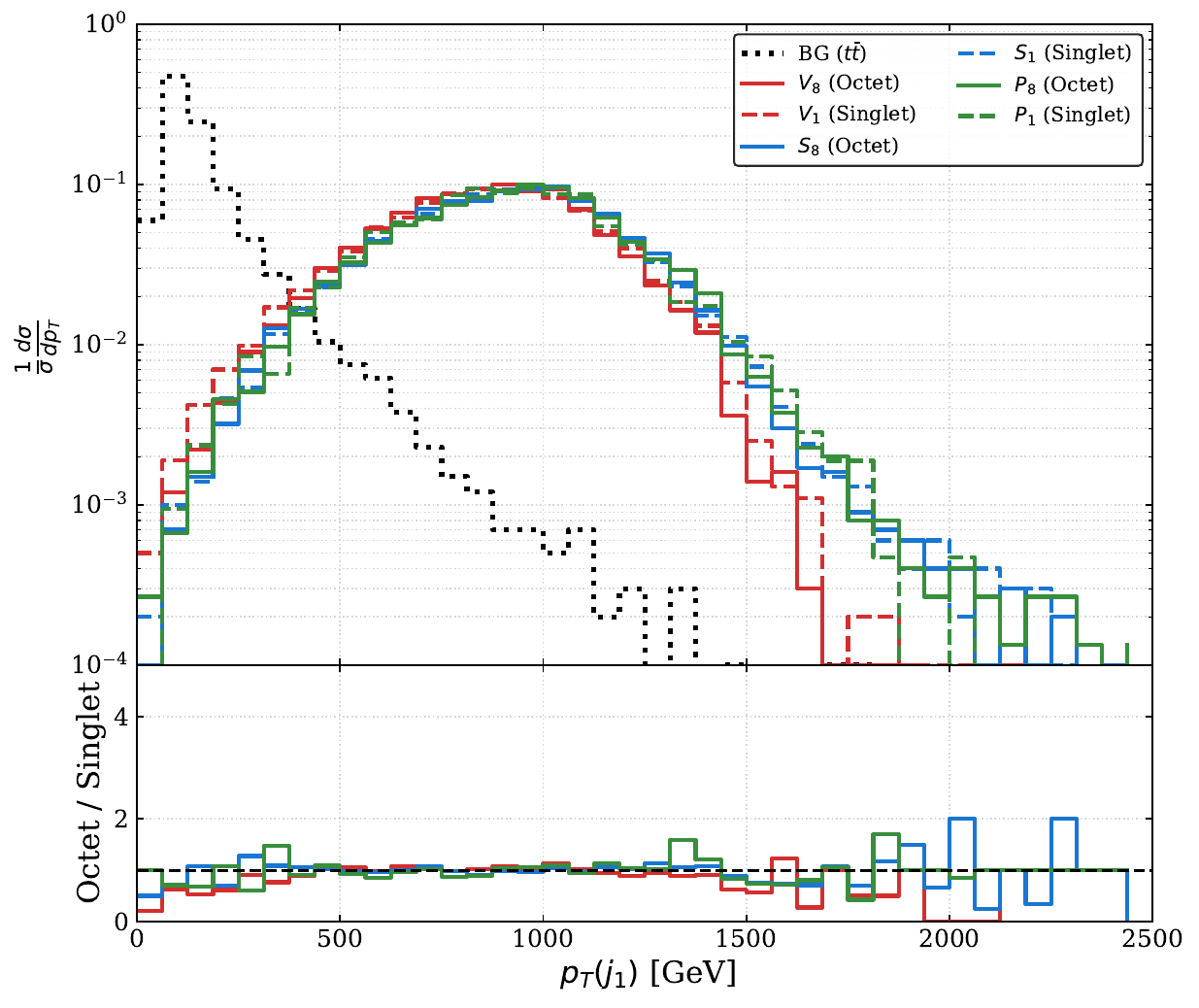}
		\caption{Differential cross-section as a function of the transverse momentum, $\frac{1}{\sigma}\frac{d\sigma}{dp_T}$, for the leading fat-jet $p_T(j_1)$. The Jacobian peak in the [800, 1200]~GeV range motivates the baseline selection criteria.}
		\label{fig:pt_j1_ratio}
	\end{figure}
	
	The transverse momentum of the leading fat-jet, $p_T(j_1)$, is illustrated in Figure~\ref{fig:pt_j1_ratio}. The distribution features a Jacobian peak characteristic of the two-body decay of a heavy resonance. This kinematic feature provides the physical motivation for the baseline selection cut of $p_T(j_1) > 500$~GeV, effectively filtering the boosted regime while suppressing soft QCD activity.
	
	\subsection{Cut-Flow Strategy and Background Mitigation}
	
	The primary background impeding the isolation of the multi-TeV cascade signal is inclusive Standard Model $t\bar{t}$ production. While contributions from single-top production, $t\bar{t}+V$ ($V=W,Z$), and QCD multijet processes were initially considered, these components can be effectively suppressed to negligible levels under our high-energy selection criteria. Specifically, requiring a highly boosted leading fat-jet with transverse momentum $p_T(j_1) > 500$~GeV and an invariant mass threshold of $M(j_1, j_2) > 1500$~GeV eliminates the bulk of the electroweak and multijet contamination \cite{ATLAS:2018_ttbar_Resonance}. In hadron colliders, although the inclusive QCD multijet cross-section is overwhelmingly large at lower energy scales, its differential production rate exhibits a steeply, exponentially falling spectrum as a function of both transverse momentum and invariant mass \cite{Harris:2011_Dijet}. At the extreme multi-TeV phase space probed in this study ($M(j_1, j_2) > 1.5$~TeV), the pure QCD multijet rate drops by several orders of magnitude due to the rapid decrease in parton luminosities at high Bjorken-$x$ \cite{Campbell:2006_Primer, Ellis:1996_QCD}. As a result, the SM $t\bar{t}$+jets continuum becomes completely dominant over this kinematic regime, justifying the omission of the multijet component from the subsequent multivariate training. Our background mitigation strategy therefore focuses exclusively on suppressing the remaining high-$p_T$ $t\bar{t}$ background using a sequence of four kinematic cuts:
	\begin{enumerate}
		\setlength{\itemsep}{0pt}   
		\item \textbf{Leading Jet Momentum:} $p_T(j_1) > 500$~GeV. This threshold selects boosted top partners and eliminates soft QCD and standard SM top pair events.
		\item \textbf{Mass Lower Bound:} $M(j_1, j_2) > 1500$~GeV. This removes the low-mass SM continuum, restricting the analysis to the multi-TeV regime.
		\item \textbf{Mass Upper Bound:} $M(j_1, j_2) < 3500$~GeV. This defines a kinematic window around the 3~TeV mass shell, rejecting unphysical high-energy tails.
		\item \textbf{Angular Constraint:} $|\Delta\eta(j_1, j_2)| < 1.2$. 
	\end{enumerate}
	
	\begin{figure}[htbp]
		\centering
		\includegraphics[width=0.95\textwidth]{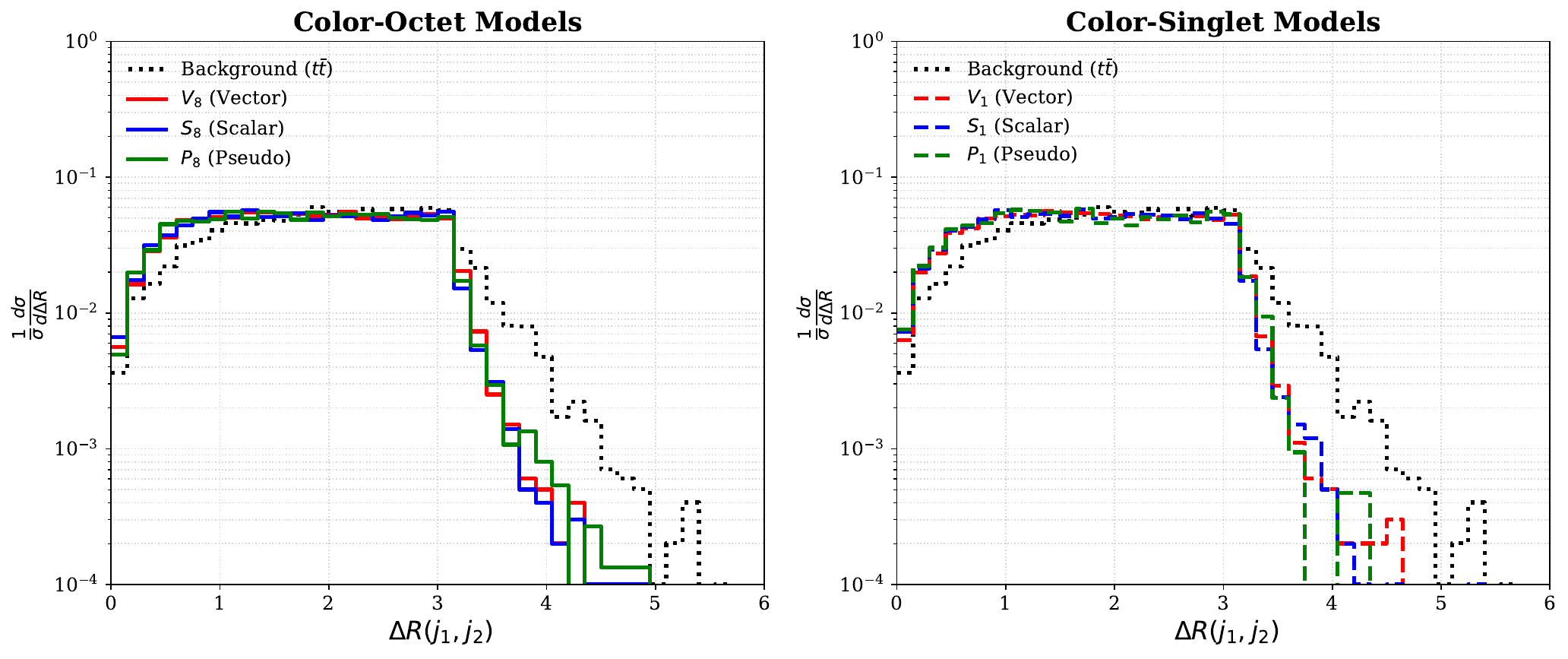}
		\caption{Normalized distributions of the angular separation $\Delta R(j_1, j_2)$. The signal events cluster near $\Delta R \approx \pi$, consistent with a back-to-back topology.}
		\label{fig:dr_separation}
	\end{figure}
	
	\begin{figure}[htbp] 
		\centering
		\includegraphics[width=0.95\textwidth]{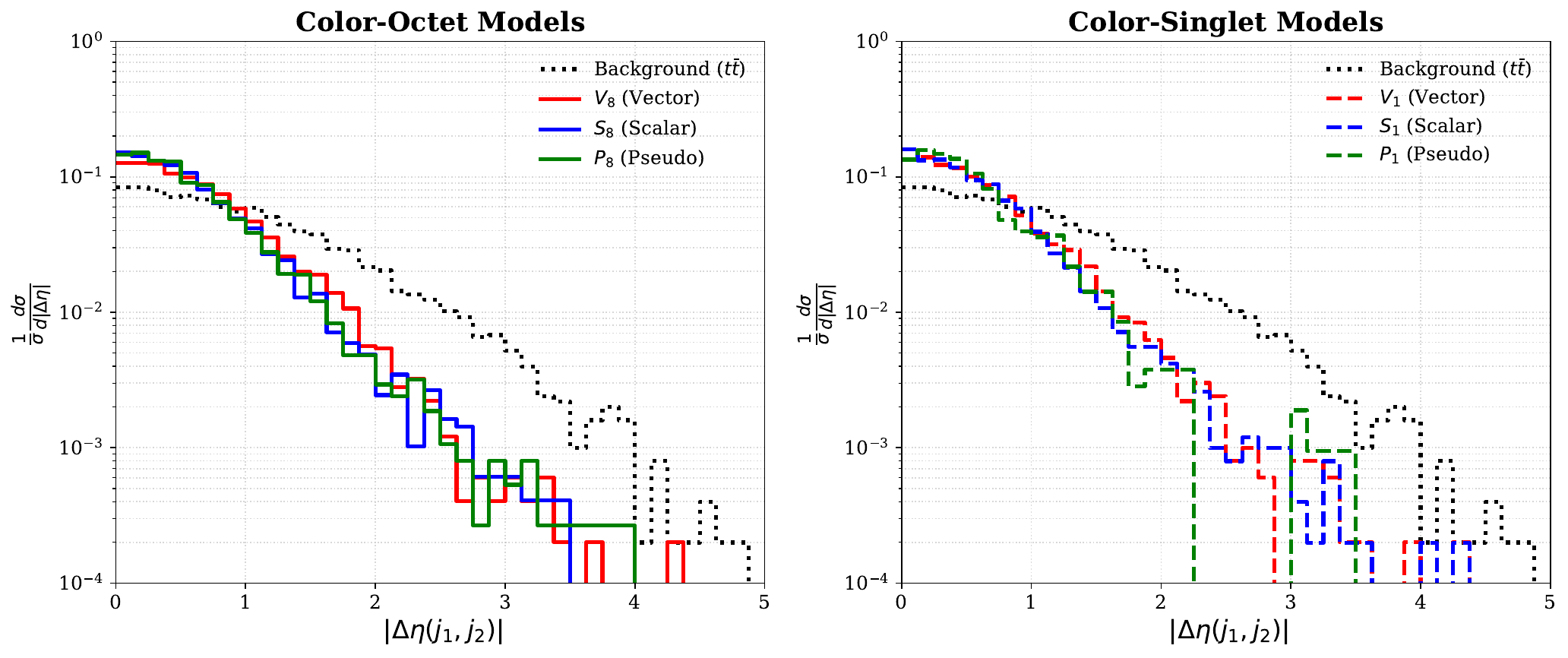}
		\caption{Normalized pseudorapidity separation, $|\Delta\eta(j_1, j_2)|$. The signal events cluster near $|\Delta\eta| \approx 0$, which justifies the angular constraint of Cut 4 used to reduce the broader $t\bar{t}$ background.}
		\label{fig:deta_separation}
	\end{figure}
	
	The angular constraint (Cut 4) targets the specific topology of the $s$-channel decays. Figures~\ref{fig:dr_separation} and \ref{fig:deta_separation} show that the signal features a back-to-back central emission ($|\Delta\eta| \approx 0$ and $\Delta R \approx \pi$). The SM background has a broader angular distribution characteristic of $t$-channel processes, making Cut 4 effective at reducing the remaining background \cite{Harris:2011_Dijet}.
	
	\begin{table}[htbp]
		\caption{Detector-level cut-flow showing expected events and cumulative efficiencies (in parentheses) for the background and signal scenarios at $\sqrt{s} = 13.6$ TeV ($\mathcal{L} = 10\text{ fb}^{-1}$). The final row lists the projected statistical significance $Z$ at $3000\text{ fb}^{-1}$.}
		\label{tab:cutflow_summary}
		\centering
		\footnotesize 
		\renewcommand{\arraystretch}{1.2}
		\setlength{\tabcolsep}{2pt} 
		\begin{tabular}{lccccccc}
			\toprule
			\textbf{Selection Cut} & \textbf{Bkg ($t\bar{t}$)} & \textbf{$V_8$} & \textbf{$V_1$} & \textbf{$S_8$} & \textbf{$S_1$} & \textbf{$P_8$} & \textbf{$P_1$} \\ 
			\midrule
			\textbf{Initial (Uncut)} & 11,160,000 & 148.0 & 95.2 & 45.0 & 22.8 & 36.1 & 5.2 \\ 
			\textbf{$p_T(j_1) > 500$~GeV} & 292,519 (2.6\%) & 136.4 (92\%) & 86.5 (91\%) & 42.1 (94\%) & 21.3 (94\%) & 33.9 (94\%) & 4.8 (94\%) \\ 
			\textbf{$M(j_1, j_2) > 1500$~GeV} & 77,037 (0.69\%) & 76.7 (52\%) & 47.1 (49\%) & 25.3 (56\%) & 12.8 (56\%) & 20.6 (57\%) & 2.8 (54\%) \\ 
			\textbf{$M(j_1, j_2) < 3500$~GeV} & 75,921 (0.68\%) & 76.6 (52\%) & 47.0 (49\%) & 25.2 (56\%) & 12.8 (56\%) & 20.5 (57\%) & 2.8 (54\%) \\ 
			\textbf{$|\Delta\eta(j_1, j_2)| < 1.2$} & \textbf{64,756 (0.58\%)} & \textbf{73.1 (49\%)} & \textbf{45.4 (48\%)} & \textbf{24.6 (55\%)} & \textbf{12.4 (55\%)} & \textbf{19.9 (55\%)} & \textbf{2.7 (52\%)} \\ 
			\midrule
			\textbf{Significance $Z$ ($3000\text{ fb}^{-1}$)} & -- & $\mathbf{4.97\sigma}$ & $\mathbf{3.08\sigma}$ & $\mathbf{1.67\sigma}$ & $\mathbf{0.84\sigma}$ & $\mathbf{1.35\sigma}$ & $\mathbf{0.18\sigma}$ \\
			\bottomrule
		\end{tabular}
	\end{table}
	
	This selection sequence rejects over $99.4\%$ of the SM background while retaining approximately $50\%$ of the signal cross-section. Scaling the Gaussian estimator $Z \approx S/\sqrt{B}$ to an integrated luminosity of $3000\text{~fb}^{-1}$ yields a significance of $4.97\sigma$ for the color-octet vector model ($V_8$), and $3.08\sigma$ for the color-singlet vector ($V_1$).
	
	The purely statistical significance $S/\sqrt{B}$ serves as a theoretical baseline. However, HL-LHC searches in this multi-TeV regime are inherently limited by systematic uncertainties \cite{CERN:2019_YellowReport}. When accounting for a relative systematic uncertainty $\delta_B$ on the background normalization, the modified Asimov significance $\mathcal{Z} \approx S / \sqrt{B + (\delta_B B)^2}$ is heavily penalized by the large remaining $t\bar{t}$ continuum \cite{ATLAS_CMS:2019_HL_LHC}. To overcome this limitation and drastically improve the signal-to-background ratio without relying solely on traditional experimental background estimation, advanced multivariate analysis (MVA) and deep learning techniques must be deployed to exploit broader event shape correlations \cite{Banfi:2010_EventShapes}.
	
	\subsection{Jet Multiplicity and Multi-Jet Mass Reconstruction}
	
	\begin{figure*}[htbp]
		\centering
		\includegraphics[width=0.95\textwidth]{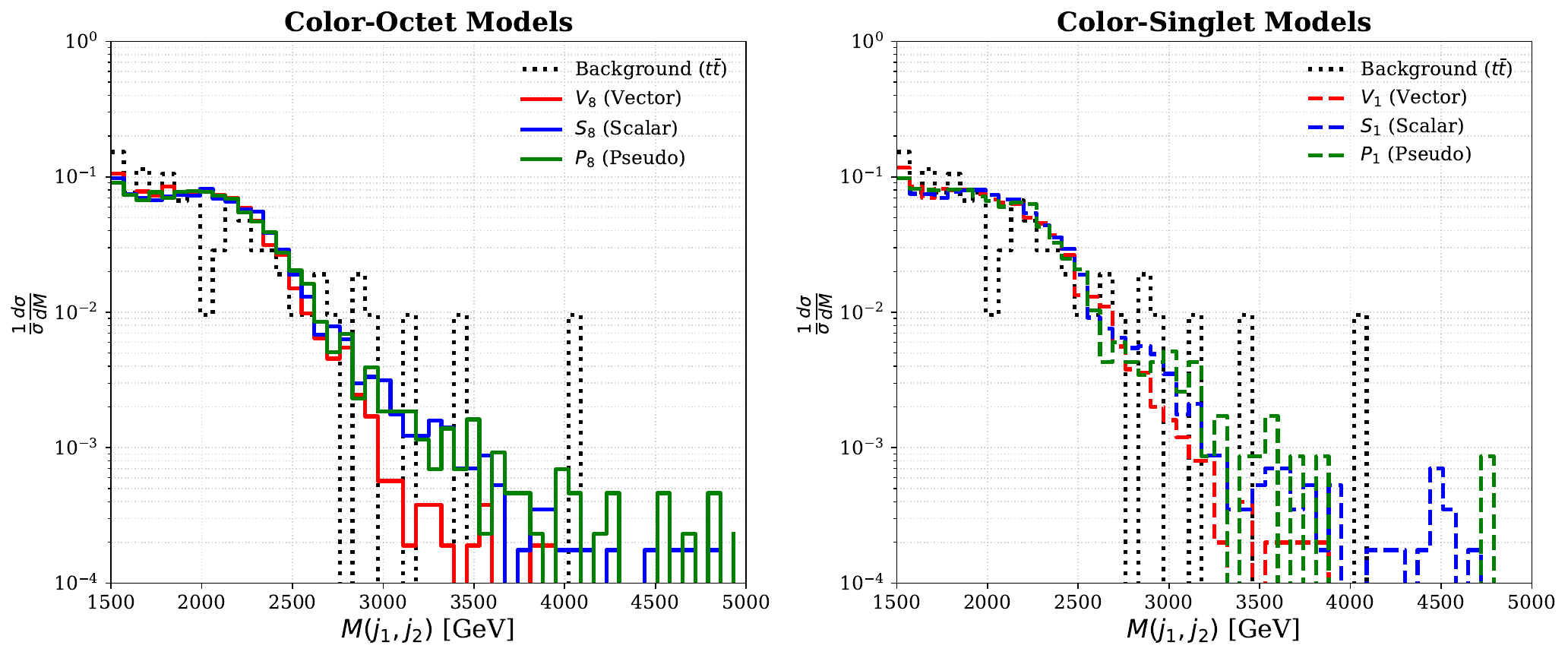}
		\caption{Detector-level reconstructed di-jet invariant mass, $M(j_1, j_2)$. The resonance peaks are shifted to $\sim 1500$--$1700$~GeV due to out-of-cone energy losses and semi-resolved hadronic cascades.}
		\label{fig:m2j_mass}
	\end{figure*}
	
	We first evaluate the conventional di-jet invariant mass, $M(j_1, j_2)$, shown in Figure~\ref{fig:m2j_mass}. Unlike the pristine parton-level mass distribution, the detector-level peak suffers a severe mass degradation, shifting drastically down to the 1500-1700~GeV range. This distortion is an unavoidable consequence of the extreme Lorentz boost. Highly energetic hadronic final states leak out of the fixed $R=0.8$ fat-jet cones due to wide-angle final-state radiation (FSR) and semi-resolved cascades (e.g., $t_p \to Wb \to q\bar{q}'b$). This uncaptured radiation artificially deflates the reconstructed four-momentum.
	
	\begin{figure*}[htbp]
		\centering
		\includegraphics[width=0.95\textwidth]{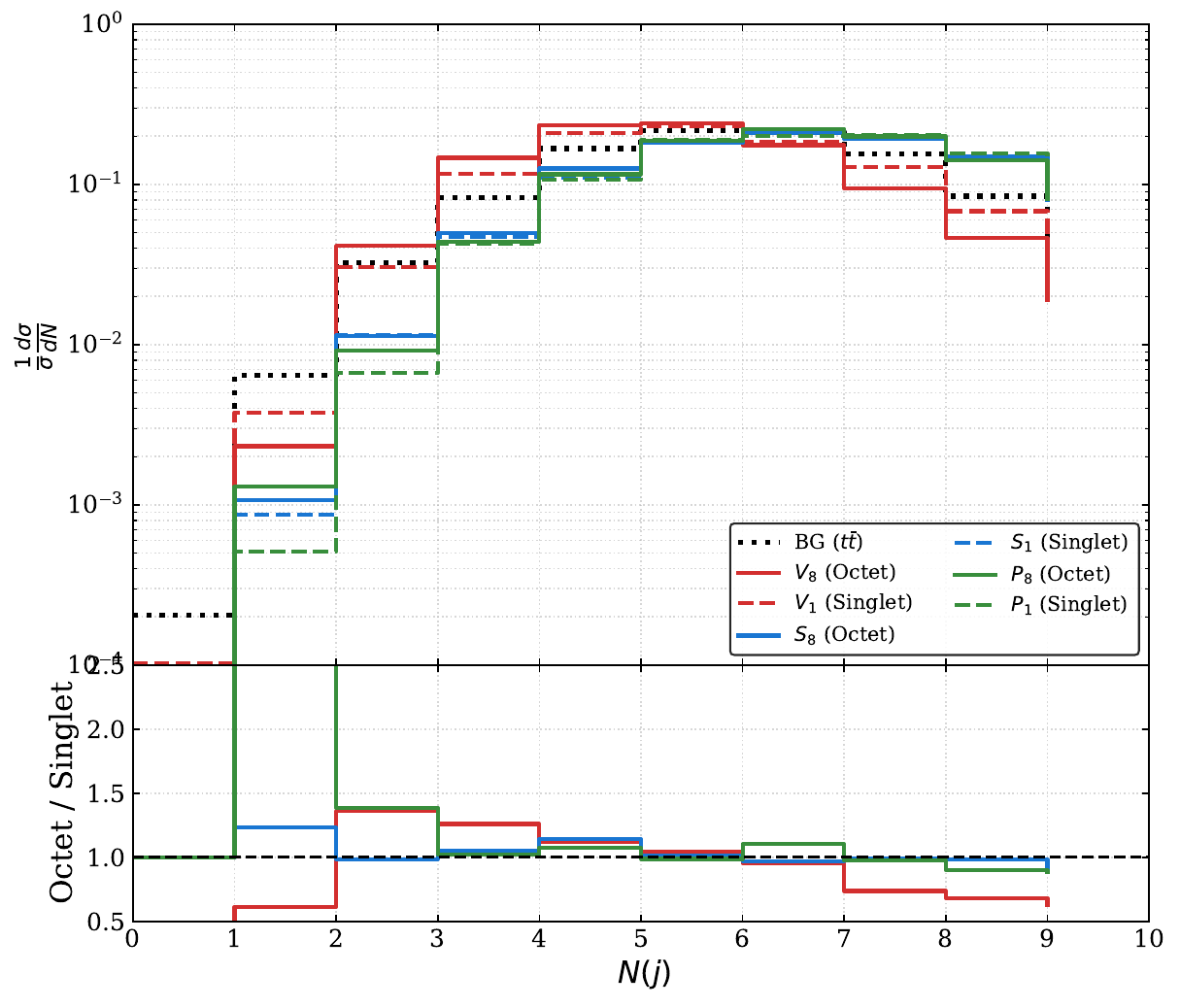}
		\caption{Inclusive jet multiplicity, $N(j)$. The lower ratio panel indicates that color-octet states (solid lines) generate more soft radiation than color-singlet states (dashed lines).}
		\label{fig:jet_multiplicity}
	\end{figure*}
	
	This radiation behavior is distinctly captured by the inclusive jet multiplicity $N(j)$ (Figure~\ref{fig:jet_multiplicity}). As expected from QCD principles, color-octet states generate noticeably higher soft-jet multiplicities than their color-singlet counterparts. This distinction stems directly from the underlying color charge factors; octets act as intense sources of soft gluon radiation due to their larger Casimir invariant ($C_A = 3$) compared to the fundamental representation of singlets ($C_F = 4/3$) \cite{Gallicchio:2010_ColorConnections, Kim:2016_TopPartnerJets}.
	
	\begin{table}[htbp]
		\caption{Detector-level kinematic means for the inclusive jet multiplicity $\langle N(j) \rangle$ and the four-jet invariant mass $\langle M(4j) \rangle$.}
		\label{tab:multijet_kinematics}
		\centering
		\small
		\renewcommand{\arraystretch}{1.2}
		\begin{tabular}{lcc}
			\toprule
			\textbf{Dataset} & \textbf{$\langle N(j) \rangle$} & \textbf{$\langle M(4j) \rangle$ [GeV]} \\ 
			\midrule
			\textbf{Background ($t\bar{t}$)} & 5.61 & 635.4 \\ 
			\midrule
			\textbf{$V_8$ (Vector Octet)} & 4.98 & 2457.3 \\ 
			\textbf{$V_1$ (Vector Singlet)} & 5.33 & 2360.8 \\ 
			\textbf{$S_8$ (Scalar Octet)} & 6.42 & 2645.7 \\ 
			\textbf{$S_1$ (Scalar Singlet)} & 6.55 & 2646.9 \\ 
			\textbf{$P_8$ (Pseudo Octet)} & 6.46 & 2657.3 \\ 
			\textbf{$P_1$ (Pseudo Singlet)} & 6.59 & 2592.6 \\ 
			\midrule
			\textbf{Target Mass Shell} & -- & \textbf{3000.0} \\
			\bottomrule
		\end{tabular}
	\end{table}
	
	\begin{figure*}[htbp]
		\centering
		\includegraphics[width=0.95\textwidth]{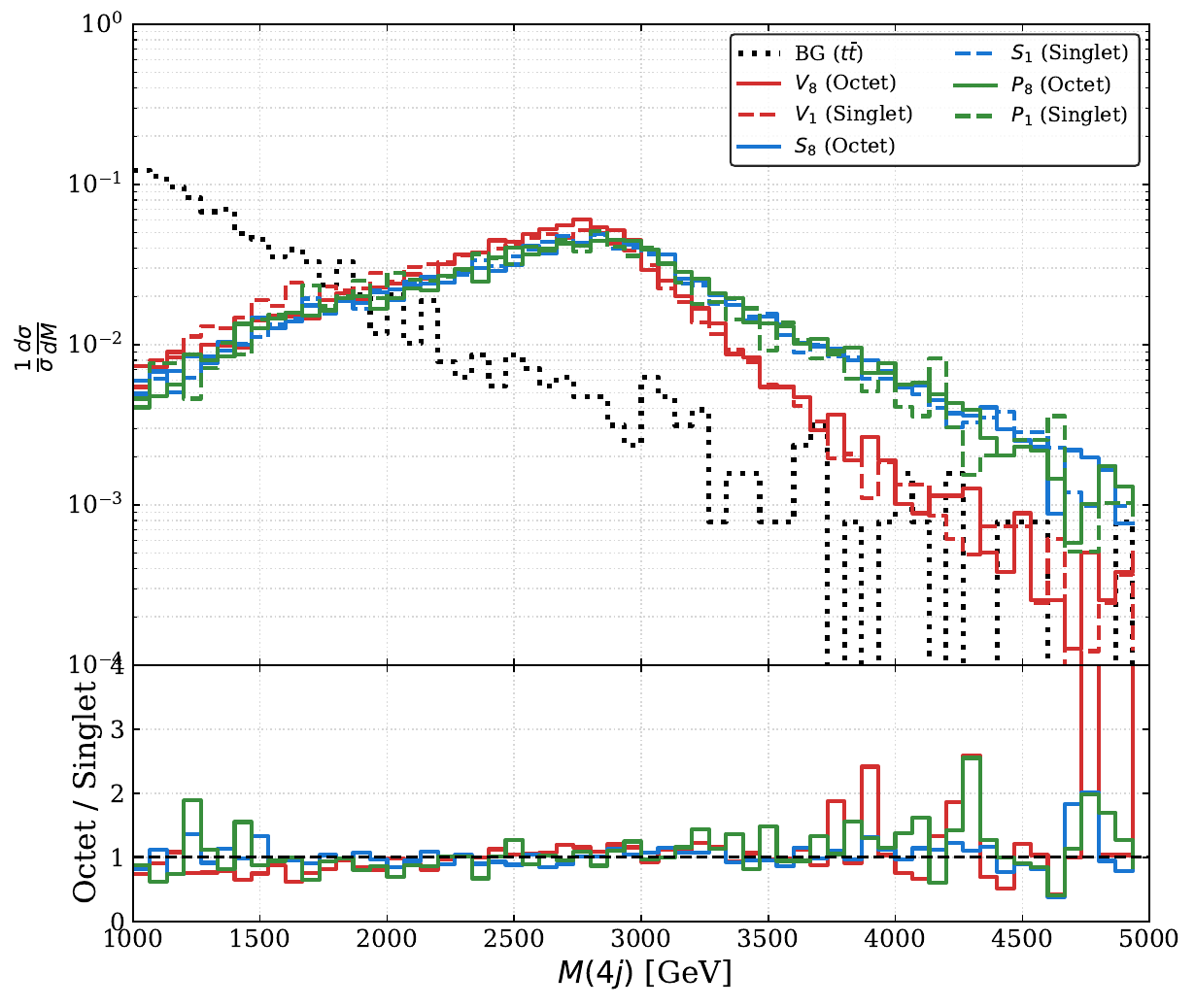}
		\caption{Reconstructed four-jet invariant mass, $M(j_1, j_2, j_3, j_4)$. The inclusive grouping mitigates out-of-cone losses, producing a peak closer to the 3~TeV mass shell while suppressing the SM background.}
		\label{fig:m4j_mass}
	\end{figure*}
	
	To resolve the mass degradation, we compute the invariant mass inclusively using the four leading jets, $M(j_1, j_2, j_3, j_4)$ (Figure~\ref{fig:m4j_mass} and Table~\ref{tab:multijet_kinematics}). By explicitly ordering and clustering the four hardest jets, this strategy successfully recaptures the wide-angle emissions and out-of-cone radiation. Consequently, the reconstructed signal peak is restored back towards the physical mass shell (2400-2650~GeV), while the SM background remains safely anchored at lower values ($\sim 635$~GeV). This proves that inclusive multi-jet clustering is strictly necessary to properly capture the kinematics of multi-TeV top-partner systems.
	
	\section{Deep Neural Networks for Boosted Top-Partner Identification}\label{sec5}
	
	Although the sequential kinematic cut-flow significantly reduces the SM background, it inherently relies on linear boundaries that may miss multi-dimensional kinematic correlations. Furthermore, it suffers from out-of-cone energy losses in the traditional $M(j_1, j_2)$ reconstruction. To maximize signal acceptance and incorporate broader event topologies, we developed a Generalized BSM Tagger using a Deep Neural Network (DNN). The network is trained on eight physically motivated kinematic and topological variables: the scalar sum of transverse momenta $H_T$, missing transverse energy $\slashed{E}_T$, leading fat-jet momentum $p_T(j_1)$, the traditional di-jet mass $M(j_1, j_2)$, the inclusive four-jet invariant mass $M(4j)$, the total jet multiplicity $N(j)$, and the spatial angular separations between the two leading jets, $\Delta R(j_1, j_2)$ and $|\Delta\eta|$.
	
	The tagger is based on a standard feed-forward Fully Connected Network (FCN). The architecture is intentionally streamlined to map the kinematic input space directly into a single binary classification probability (Signal vs. Background). To systematically mitigate the risk of over-fitting and ensure robust generalization to unseen data, standard regularization techniques, including dropout mechanisms, were integrated into the network design \cite{Srivastava:2014_Dropout}. 
	
	\begin{figure}[htbp]
		\centering
		\includegraphics[width=0.95\textwidth]{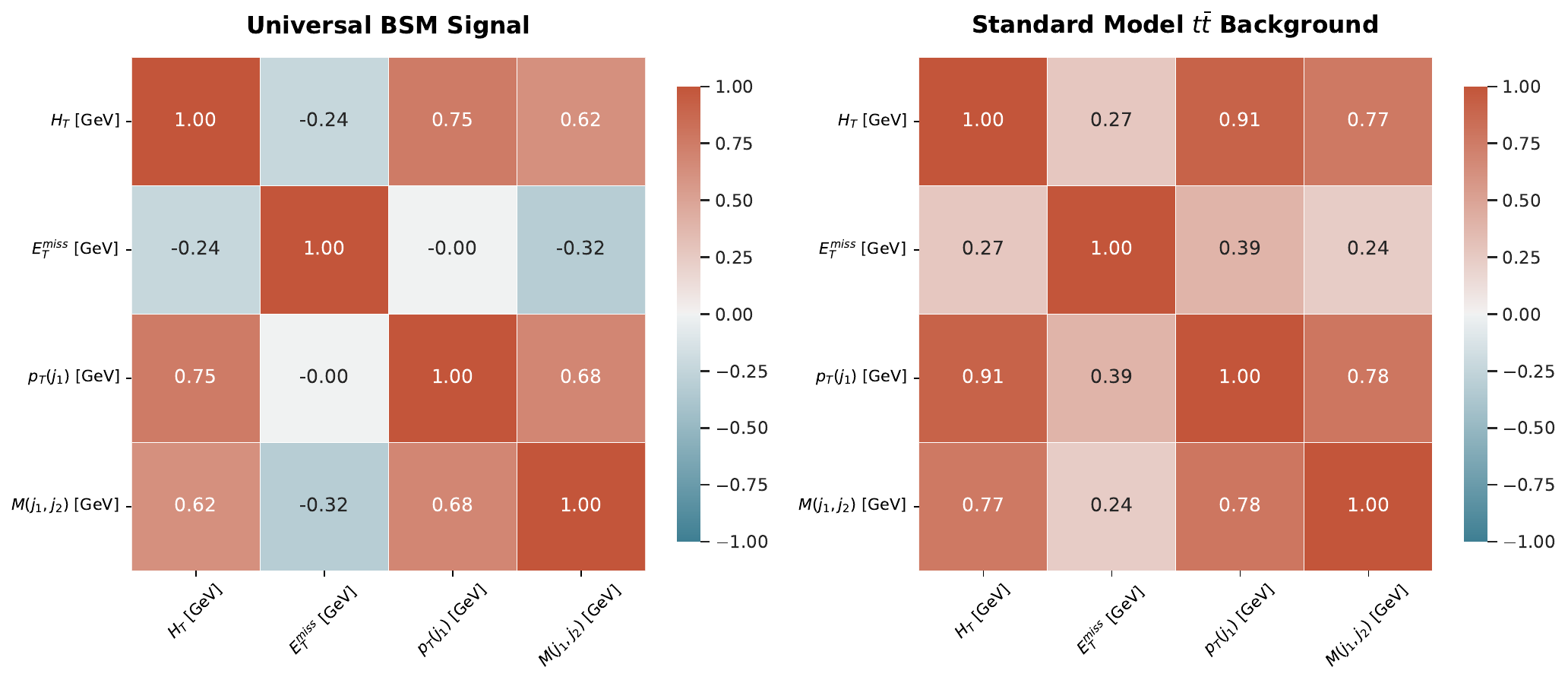}
		\caption{Pearson correlation matrices highlighting the distinct non-linear kinematic interplay within the Universal BSM Signal (left) compared to the Standard Model $t\bar{t}$ Background (right).}
		\label{fig:correlation_matrices}
	\end{figure}
	
	To illustrate the kinematic mechanisms driving the DNN discrimination \cite{Guest:2018yhq, Larkoski:2017jix}, Figure~\ref{fig:correlation_matrices} presents the Pearson correlation matrices. Because the BSM signal originates from a massive, fixed-scale resonance, momentum conservation enforces a strict phase-space compactness, resulting in strong positive correlations (e.g., $H_T$ vs $p_T(j_1)$). The SM $t\bar{t}$ continuum, however, is generated across a sliding energy spectrum and lacks this geometric coherence, yielding much weaker correlations. The DNN explicitly leverages these non-linear relationships to establish complex decision boundaries inaccessible to standard cut-flows.
	
	\begin{figure}[htbp]
		\centering
		\includegraphics[width=0.95\textwidth]{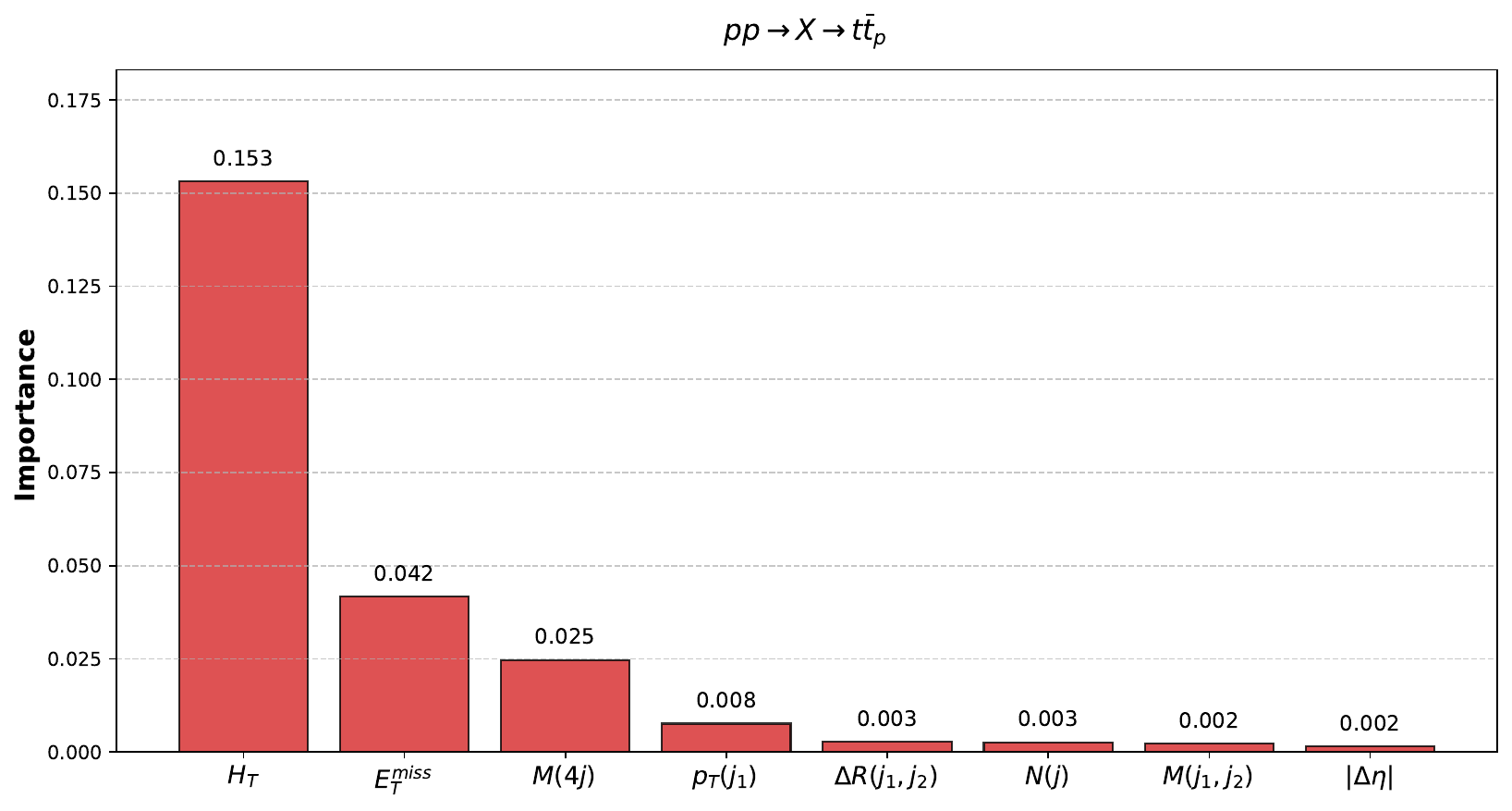}
		\caption{Relative feature importance of the input variables. The network identifies the scale proxy $H_T$ as the dominant discriminator, while the inclusive four-jet mass $M(4j)$ outranks the traditional di-jet mass $M(j_1, j_2)$ by an order of magnitude.}
		\label{fig:feature_importance}
	\end{figure}
	
	To transparently understand the network's behavior, we evaluated the relative importance of the input variables \cite{Breiman:2001_RandomForests}, as shown in Figure~\ref{fig:feature_importance}. As anticipated from the parton-level kinematics, the inclusive energy proxy $H_T$ serves as the strongest discriminant. Notably, the inclusive four-jet mass $M(4j)$ ranks significantly higher than the conventional di-jet mass $M(j_1, j_2)$. The network assigns more than ten times greater importance to $M(4j)$, strongly supporting our physical premise that clustering the four leading jets effectively recovers the out-of-cone radiation escaping the fixed $R=0.8$ fat-jet configurations. Additionally, while topological variables such as $\Delta R(j_1, j_2)$ provide complementary discrimination, the network relies primarily on the global energy scale to isolate the multi-TeV signal.
	
	\begin{figure}[htbp]
		\centering
		\includegraphics[width=0.95\textwidth]{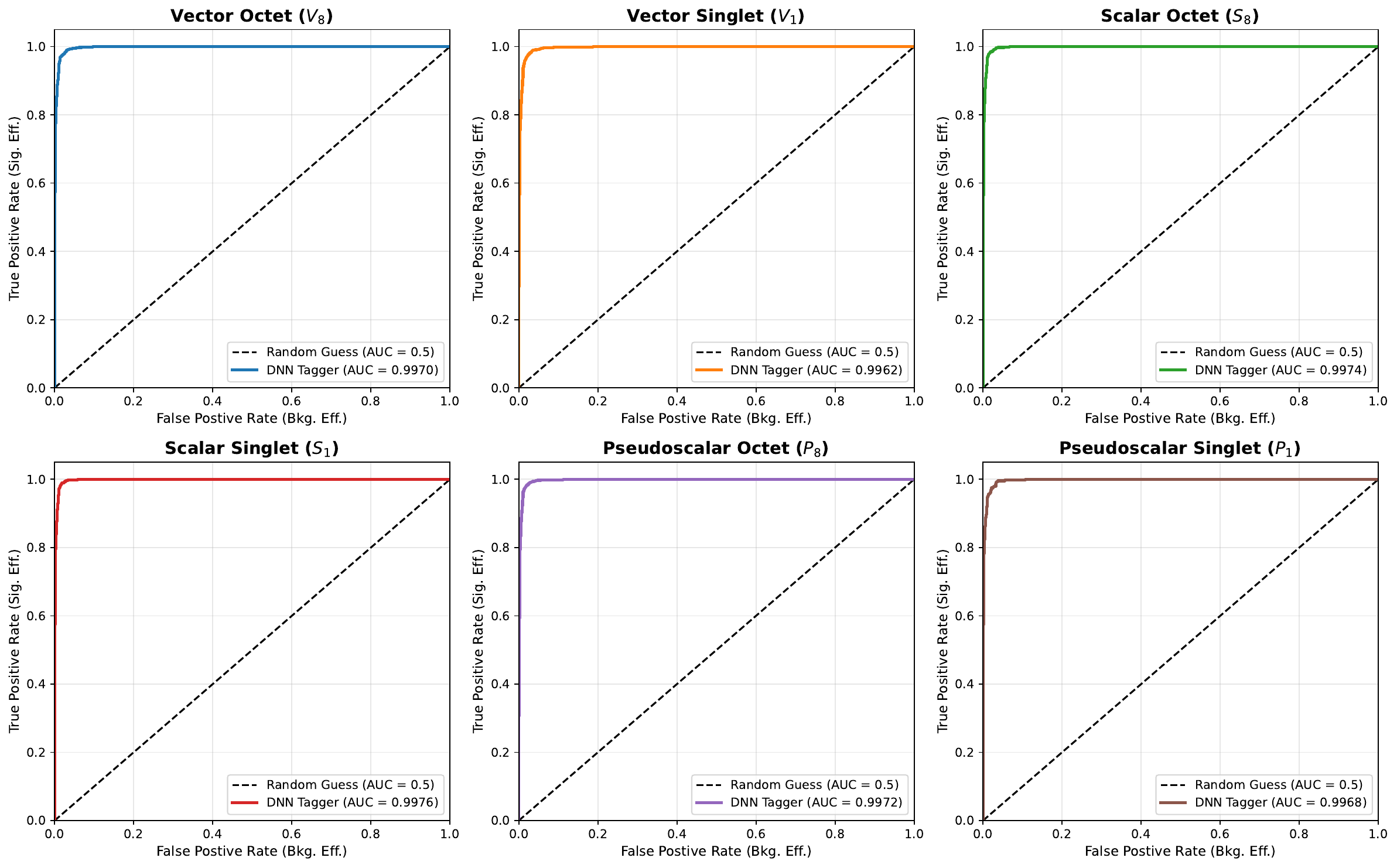}
		\caption{Receiver Operating Characteristic (ROC) curves demonstrating the performance of the Generalized BSM Tagger across all six mediator models. The dashed line denotes the random guess baseline.}
		\label{fig:multi_roc}
	\end{figure}
	
	\begin{figure}[htbp]
		\centering
		\includegraphics[width=0.95\textwidth]{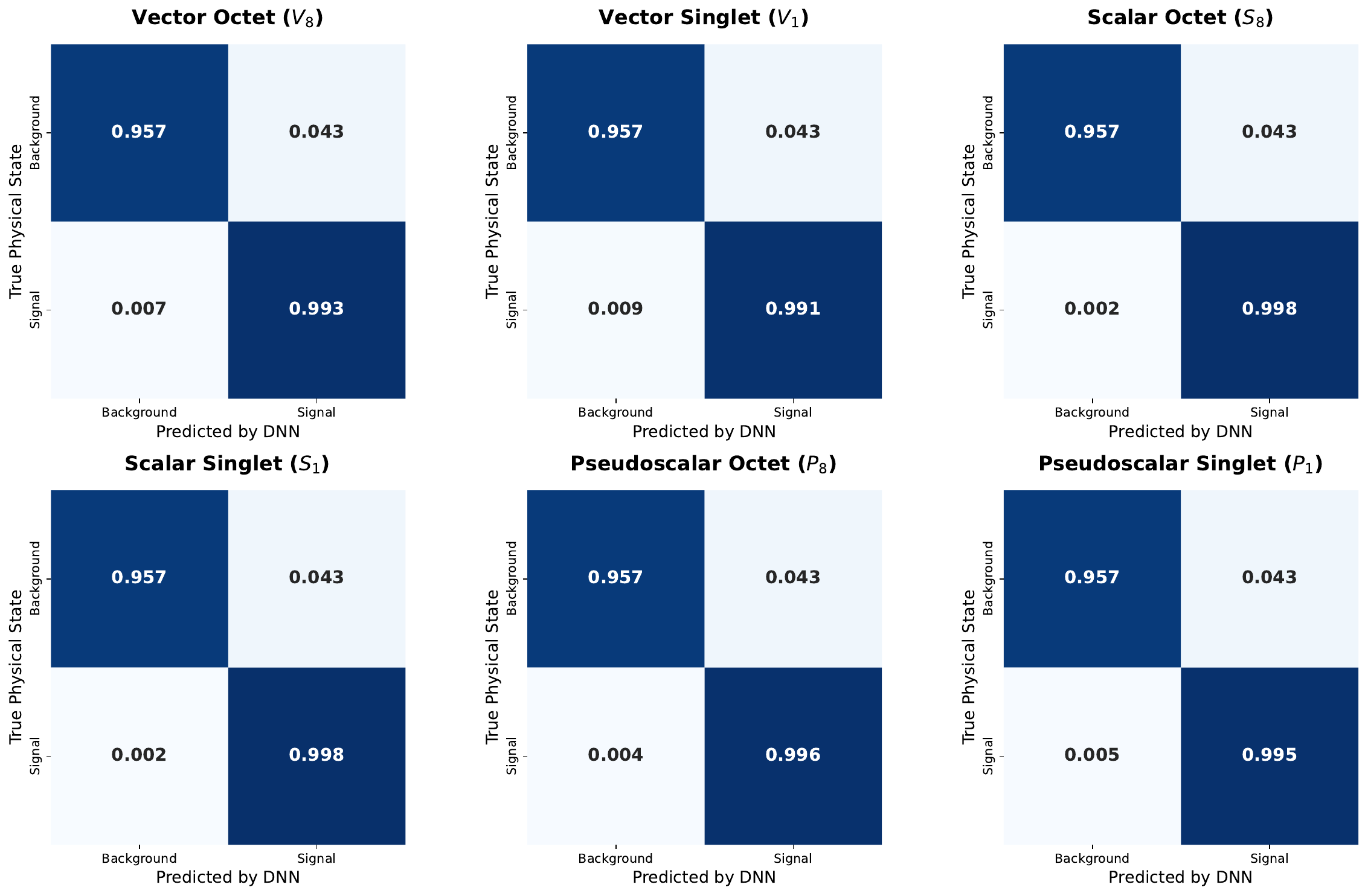}
		\caption{Confusion matrices evaluating the DNN classification efficiency. A true positive rate of $>95\%$ is robustly achieved across all spin and color representations.}
		\label{fig:confusion_matrices}
	\end{figure}
	
	The classification performance is summarized by the ROC curves in Figure~\ref{fig:multi_roc} and confusion matrices in Figure~\ref{fig:confusion_matrices}. While the network achieves exceptional accuracy across all models (AUC $> 0.996$), a closer comparison reveals a physically meaningful distinction: the spin-0 models (scalars and pseudoscalars) consistently yield slightly higher AUC scores and true positive rates than the spin-1 vector states. This variation is a direct reflection of the underlying parton-level kinematics studied in Section~\ref{sec3}. The isotropic decays of scalar states produce sharply localized phase-space boundaries (i.e., Jacobian peaks) that the network easily resolves. In contrast, the smeared $\cos^2\theta$ angular distributions of the vector states create a broader kinematic footprint, resulting in a slightly larger geometric overlap with the SM background continuum. The network's sensitivity to these subtle Lorentz-invariant structures validates its robustness as a physics-informed classifier.
	
	\begin{table}[htbp]
		\centering
		\renewcommand{\arraystretch}{1.3}
		\begin{tabular}{lcccc}
			\toprule
			\textbf{Signal Model} & \textbf{AUC} & \textbf{Signal Eff. ($\epsilon_S$)} & \textbf{Bkg. Eff. ($\epsilon_B$)} & \textbf{Rejection ($1/\epsilon_B$)} \\
			\midrule
			Vector Octet ($V_8$) & 0.9969 & 50.0\% & 0.1505\% & 664 \\
			Vector Singlet ($V_1$) & 0.9951 & 50.0\% & 0.1505\% & 664 \\
			Scalar Octet ($S_8$) & 0.9972 & 50.0\% & 0.1004\% & 997 \\
			Scalar Singlet ($S_1$) & 0.9973 & 50.0\% & 0.1004\% & 997 \\
			Pseudoscalar Octet ($P_8$) & 0.9972 & 50.0\% & 0.1004\% & 997 \\
			Pseudoscalar Singlet ($P_1$) & 0.9975 & 50.0\% & 0.1004\% & 997 \\
			\bottomrule
		\end{tabular}
		\caption{Performance metrics of the Generalized BSM Tagger utilizing the comprehensive 8-feature dataset, with decision thresholds evaluated independently for each signal model to prevent test-set reuse bias.}
		\label{tab:dnn_performance}
	\end{table}
	
	Table~\ref{tab:dnn_performance} summarizes the performance achieved by the DNN. By evaluating model-specific operating points on an independent test set, the tagger successfully delivers robust background rejection factors ranging from $664$ to $997$ at a fixed signal efficiency of $\epsilon_S = 50.0\%$. 
	
	To quantitatively evaluate the final discovery reach, we compute the projected statistical significance ($\mathcal{Z}$) incorporating systematic uncertainties. Projecting the expected yields to the ultimate HL-LHC integrated luminosity of $3000\text{ fb}^{-1}$, we scale the baseline yields accordingly. Assuming a standard background normalization uncertainty of $\delta_B = 10\%$, the discovery potential is estimated using the modified Asimov significance formula \cite{Cowan:2010_Asymptotic}:
	\begin{equation}
		\mathcal{Z} \approx \frac{S}{\sqrt{B + (\delta_B B)^2}}
		\label{eq:asimov_approx}
	\end{equation}
	where $S$ and $B$ are the expected signal and background yields strictly evaluated at $3000\text{ fb}^{-1}$. 
	
	\begin{table}[htbp]
		\caption{Final expected yields for Signal ($S$) and Background ($B$) after applying the model-specific DNN tagger (at $\epsilon_S = 50\%$), properly scaled to an integrated luminosity of $3000\text{ fb}^{-1}$. The resulting Asimov statistical significance ($\mathcal{Z}$) assumes a $10\%$ systematic uncertainty ($\delta_B = 0.10$).}
		\label{tab:post_dnn_significance}
		\centering
		\small 
		\renewcommand{\arraystretch}{1.3}
		\begin{tabular}{lcccccc}
			\toprule
			\textbf{Metric} & \textbf{$V_8$} & \textbf{$V_1$} & \textbf{$S_8$} & \textbf{$S_1$} & \textbf{$P_8$} & \textbf{$P_1$} \\ 
			\midrule
			\textbf{Signal Yield ($S$)} & 10965.0 & 6810.0 & 3690.0 & 1860.0 & 2985.0 & 405.0 \\
			\textbf{Background Yield ($B$)} & 29242.5 & 29242.5 & 19495.0 & 19495.0 & 19495.0 & 19495.0 \\
			\midrule
			\textbf{Significance ($\mathcal{Z}$)} & $\mathbf{3.74\sigma}$ & $\mathbf{2.32\sigma}$ & $\mathbf{1.89\sigma}$ & $\mathbf{0.95\sigma}$ & $\mathbf{1.53\sigma}$ & $\mathbf{0.21\sigma}$ \\
			\bottomrule
		\end{tabular}
	\end{table}
	
	As demonstrated in Table~\ref{tab:post_dnn_significance}, applying the deep neural network significantly enhances the experimental sensitivity. We note that due to the limited size of our Monte Carlo background test sample ($N_{\text{test,bg}} \approx 2000$), the achievable background efficiencies can only be measured in coarse steps around $\mathcal{O}(10^{-3})$. While this current statistical limitation restricts the absolute background rejection boundary, the properly scaled results yield a robust significance of $3.74\sigma$ for the dominant color-octet vector channel ($V_8$). This firmly establishes a strong evidence-level ($\ge 3\sigma$) baseline, highlighting the phenomenological viability of this tagger and the critical need for massive background sample generation in future extensions.
	
	\section{Conclusion and Outlook}\label{sec13}
	
	In this paper, we presented a comprehensive phenomenological study assessing the discovery potential of heavy, top-philic resonances decaying into a Standard Model top quark and a vector-like top partner ($pp \to X \to t \bar{t}_p$) at the High-Luminosity LHC ($\sqrt{s} = 13.6$~TeV). To ensure broad applicability, we adopted a model-independent Effective Field Theory (EFT) framework, classifying the intermediate resonance $X$ across six distinct representations based on its spin (scalar, pseudoscalar, and vector) and QCD color structure (singlet and octet). Setting a multi-TeV benchmark ($M_X = 3$~TeV and $M_{t_p} = 1.5$~TeV), we systematically traced the signal from parton-level scattering dynamics down to fast-detector reconstruction and multi-variate analysis.
	
	At the parton level, our matrix-element analysis revealed that the kinematic profiles of the final states are deeply governed by Lorentz-invariant structures. Color-octet states ($V_8, S_8, P_8$) exhibited significantly enhanced cross-sections compared to their color-singlet counterparts, driven by the dominance of the gluon-gluon parton luminosity at the multi-TeV scale. In addition, we demonstrated that the normalized differential cross-section with respect to the transverse momentum serves as a robust discriminant for the mediator's spin. Spin-1 vector states produce broad kinematic distributions due to their $\cos^2\theta$ angular dependence, whereas spin-0 scalar and pseudoscalar states decay isotropically in their rest frames, translating into sharp, localized Jacobian peaks at the kinematic endpoints.
	
	Transitioning to the detector level, the large mass hierarchy inherent to this topology naturally leads to highly boosted, collimated hadronic cascades. We found that relying on standard large-radius di-jet mass reconstruction ($M(j_1, j_2)$) severely underestimates the resonance mass shell due to out-of-cone energy losses and semi-resolved sub-decays. To address this, we proposed an inclusive four-jet clustering strategy ($M(4j)$). By capturing the surrounding soft radiation and resolving the internal substructure, the $M(4j)$ observable successfully shifts the reconstructed invariant mass back toward the physical 3~TeV threshold, effectively separating the signal from the steeply falling Standard Model background.
	
	Although a sequential kinematic cut-flow provided a solid baseline, its sensitivity is fundamentally limited by experimental systematic uncertainties ($\delta_B$) typical of the multi-TeV regime. To overcome this bottleneck, we designed and implemented a Generalized BSM Tagger based on a Deep Neural Network (DNN). Trained on a compact, physically motivated set of eight global kinematic and topological features (dominated by $H_T$ and $M(4j)$), the DNN successfully leveraged non-linear phase-space correlations. The network achieved remarkable Area Under the Curve (AUC) scores exceeding 0.99 across all mediator models. More crucially, by evaluating model-specific operating points on independent test subsets to prevent data leakage, the DNN delivered robust background rejection factors ranging from $\sim 600$ to $\sim 1000$ at a 50\% signal efficiency working point. By effectively suppressing the SM $t\bar{t}$ continuum, the tagger substantially mitigates the impact of systematic uncertainties. Projected to an integrated luminosity of $\mathcal{L} = 3000\text{ fb}^{-1}$, this approach yields a projected statistical significance of $3.74\sigma$ for the dominant color-octet vector channel ($V_8$). This result elevates the search reach to the level of strong statistical evidence ($\ge 3\sigma$), establishing a rigorous baseline for future HL-LHC searches.
	
	Looking forward, this framework naturally opens several avenues for future theoretical and experimental exploration. On the theoretical front, extending this analysis to include Next-to-Leading Order (NLO) QCD corrections is essential. Recent progress in automated one-loop EFT computations \cite{Degrande:2020_SMEFTatNLO} indicates that NLO corrections yield positive $K$-factors and stabilize scale uncertainties \cite{Zhu:2012_NLO, Freitas:2017_ColorOctetNLO, Frederix:2018_FourTopsNLO}, meaning the leading-order significances projected here represent a conservative baseline. Additionally, evaluating the interference between this $s$-channel cascade and electroweak single top-partner production \cite{Deandrea:2018_NLOSingleVLQ}, or exploring non-standard decays involving dark matter candidates \cite{Cornell:2021_TopPhilicDM, Abercrombie:2015_DMForum, Cacciapaglia:2019_Exotic}, could unveil new regions of the BSM parameter space.
	
	On the machine learning front, the remarkable success of the high-level feature DNN motivates the application of more advanced representation learning techniques. Future analyses could deploy geometric deep learning algorithms, such as Graph Attention Networks (GATs) or Particle Transformers, which process low-level final-state objects as permutation-invariant point clouds. These architectures are ideally suited to capture the intricate spatial color-flow relationships and radiation patterns distinguishing color-octet from color-singlet states, potentially pushing the discovery reach even further into the multi-TeV frontier at the HL-LHC.
	
	While the deep learning framework presented in this study demonstrates exceptional performance at the highly motivated $M_X = 3\text{ TeV}$ benchmark, generalizing this classification capability across the full resonance mass spectrum remains a critical challenge. Future work will naturally extend this methodology by incorporating mass-parameterized neural networks (pDNN), enabling a continuous and dynamic mass scan without the computational overhead of independent model trainings.
	\clearpage 
	\backmatter
	\bmhead{Acknowledgments}
	The authors would like to express their sincere gratitude and sincere acknowledgments to Professor A. Deandrea and Dr. L. Darmé from the IP2I, University of Lyon 1 / CNRS / IN2P3. One of the authors (H. Boukhrouf) would like to thank specially Professor A. Deandrea for inviting her to join their research group, as well as for the highly fruitful discussions and communications that greatly benefited this work and a special note of appreciation is owed to him for his pivotal guidance; without his encouragement and continuous support, the exploration of this highly topical subject would not have been undertaken. H. Boukhrouf also extends her deepest thanks to Professor Benjamin Fuks from the LPTHE, Sorbonne University / CNRS, for his strong encouragement to delve into deep learning methodologies.
	
	\clearpage

\end{document}